%% file: directed-triangles-journal-resubmission_clean.tex
\documentclass[aps,pre,reprint,amsmath,amssymb]{revtex4-1}
\usepackage{amsmath,amsfonts,amssymb,amsthm}
\usepackage[usenames]{color}
\usepackage{bm}
\usepackage{xspace}
\usepackage{paralist}
\usepackage{graphicx}
\usepackage{epstopdf}
\usepackage{boxedminipage}
\usepackage{subfig}
\usepackage[colorlinks,urlcolor=blue,citecolor=blue,linkcolor=blue]{hyperref}
\usepackage{subfig}
\usepackage{tabularx}
\usepackage{array}
\usepackage{booktabs}
\usepackage{textcase}
\usepackage{multirow}
\usepackage{braket}
\usepackage{tikz}
\usetikzlibrary{arrows,positioning,backgrounds}

\usepackage{rotating}
\usepackage{algorithm}
\usepackage{algpseudocode}
\usepackage{slashbox}
\usepackage{caption}
\newcolumntype{M}{>{$\vcenter\bgroup\hbox\bgroup}c<{\egroup\egroup$}}
\DeclareCaptionType{copyrightbox}

\newcommand{\SecLabel}[1]{\hyperref[sec:#1]{\textbf{Section \ref*{sec:#1}}.}} %section
\newcommand{\Sec}[1]{\hyperref[sec:#1]{\S\ref*{sec:#1}}} %section
\newcommand{\Eqn}[1]{\hyperref[eq:#1]{(\ref*{eq:#1})}} %equation
\newcommand{\Fig}[1]{\hyperref[fig:#1]{Figure~\ref*{fig:#1}}} %figure
\newcommand{\Tab}[1]{\hyperref[tab:#1]{Table~\ref*{tab:#1}}} %table
\newcommand{\Thm}[1]{\hyperref[thm:#1]{Theorem~\ref*{thm:#1}}} %theorem
\newcommand{\Lem}[1]{\hyperref[lem:#1]{Lem.\,\ref*{lem:#1}}} %lemma
\newcommand{\Prop}[1]{\hyperref[prop:#1]{Prop.~\ref*{prop:#1}}} %property
\newcommand{\Cor}[1]{\hyperref[cor:#1]{Cor.~\ref*{cor:#1}}} %corollary
\newcommand{\Def}[1]{\hyperref[def:#1]{Defn.~\ref*{def:#1}}} %definition
\newcommand{\Alg}[1]{\hyperref[alg:#1]{Alg.~\ref*{alg:#1}}} %algorithm
\newcommand{\Ex}[1]{\hyperref[ex:#1]{Ex.~\ref*{ex:#1}}} %example
\newcommand{\Clm}[1]{\hyperref[clm:#1]{Claim~\ref*{clm:#1}}} %example
\def\linkaba{stealth'-stealth'}
\def\linkab{-stealth'}
\def\linkba{stealth'-}

\newcommand{\directedtriangle}[5]{
  \node (A) at (0,0) [nd] {}; % Lower Left
  \node (B) at +(0.7,0) [nd] {}; % Lower Right
  \node (C) at +(0.35,0.6) [nd] {}; % Top
  \draw[#1] (A)--(B);
  \draw[#2] (A)--(C);
  \draw[#3] (B)--(C);
  \node [inner sep = 0,below] at +(0.35,-0.2) {\footnotesize #4 #5};
  \node [inner sep = 0,below] at +(0.35,-0.2) {\phantom{\footnotesize (vi) reciprocal}};
}
\newcommand{\dwedge}[3]{
\begin{tikzpicture}[nd/.style={circle,draw,fill=teal!50,inner sep=2pt}]      
  \node (A) at (0,0) [nd] {}; % Lower Left
  \node (B) at +(0.7,0) [nd] {}; % Lower Right
  \node (C) at +(0.35,0.6) [nd] {}; % Top
  \draw[#1] (A)--(C);
  \draw[#2] (B)--(C);
\node at (0.35,0.85) {$v$};   \end{tikzpicture}
}
\newcommand{\directedwedge}[4]{
  \node (A) at (0,0) [nd] {}; % Lower Left
  \node (B) at +(0.7,0) [nd] {}; % Lower Right
  \node (C) at +(0.35,0.6) [nd] {}; % Top
  \draw[#1] (A)--(C);
  \draw[#2] (B)--(C);
  \node [inner sep = 0,below] at +(0.35,-0.2) {\footnotesize #3 #4};
  \node [inner sep = 0,below] at +(0.35,-0.2) {\phantom{\footnotesize (vi) reciprocal}};
}
\newcommand{\wout}{out\xspace}
\newcommand{\wmid}{path\xspace}
\newcommand{\win}{in\xspace}
\newcommand{\wrecipout}{out+\xspace}
\newcommand{\wrecipin}{in+\xspace}
\newcommand{\wreciptot}{reciprocal\xspace}
\newcommand{\ttrans}{acyclic\xspace}
\newcommand{\tcycle}{cycle\xspace}
\newcommand{\toutrecip}{out+\xspace}
\newcommand{\tmidrecip}{cycle+\xspace}
\newcommand{\tinrecip}{in+\xspace}
\newcommand{\ttworecip}{cycle++\xspace}
\newcommand{\tthreerecip}{reciprocal\xspace}

\newcommand{\din}[1]{d_{#1}^{\leftarrow}}
\newcommand{\dout}[1]{d_{#1}^{\rightarrow}}
\newcommand{\drec}[1]{d_{#1}^{\leftrightarrow}}
\newcommand{\mins}{m^{\leftarrow}}
\newcommand{\mout}{m^{\rightarrow}}
\newcommand{\mrec}{m^{\leftrightarrow}}

\newcommand\EX{\mathbb{E}}

\tikzstyle{vertex}=[circle,draw=black,fill=teal!50,thin,inner sep=0pt,minimum size=6pt]

\def\typea{blue}
\def\typeb{lightgray}
\def\typec{red}
\def\typed{yellow}
\def\typee{green}
\def\typef{brown}
\def\typeg{pink}
\def\typex{lime}
\def\typey{olive}
\def\typez{teal}

\newcommand{\gcc}{\kappa}

\begin{document}
\title{Directed closure measures for networks with reciprocity}
\author{C. Seshadhri}
\email{scomand@sandia.gov}
\author{Ali Pinar}
\email{apinar@sandia.gov}
\author{Nurcan Durak}
\email{nurcan.durak@gmail.com}
\author{Tamara G. Kolda}
\email{tgkolda@sandia.gov}
\affiliation{Sandia National Laboratories, Livermore, CA, USA}
\date{\today}
\begin{abstract} The study of triangles in graphs is a standard
tool in network analysis, leading to measures such as the \emph{transitivity}, i.e.,
the fraction of paths of length $2$ that participate in triangles. 
Real-world networks are often directed, and it can be difficult to ``measure" this network structure
meaningfully.
We propose a collection of \emph{directed closure values} for measuring triangles in directed graphs
in a way that is analogous to transitivity in an undirected graph.
Our study of these values reveals much information about directed triadic closure. 
For instance, we immediately see that reciprocal edges have a high propensity to participate in triangles. 
We also observe striking similarities
between the triadic closure patterns of different web and social networks.
We perform mathematical and empirical analysis showing that directed configuration
models that preserve reciprocity cannot capture the triadic closure patterns of real networks.
\end{abstract}
\maketitle
\section{Introduction}
\label{sec:intro}
The study of triangles is by now a classic tool in the analysis of large-scale networks. The focus
on triangles has its roots in a variety of disciplines: in social sciences as a manifestation of
theories of edge formation~\cite{HoLe70,Burt04}, in physics as a local measure of clustering~\cite{WaSt98}, in biology as motifs~\cite{Milo2002}. 
Capturing triangle structure in generative models is also of great interest \cite{SeKoPi12,DuPiKo12}.
We consider the problem of studying triangles in directed networks.
Most social, communication, cyber, and  web networks
are directed networks. In directed networks, it has been observed that there is generally a significant percentage of \emph{reciprocal
edges} \cite{NeFoBa02,GaLo04,GaLo06,ZaZl+08,MiKoGuDrBh08,ZlSt09,KwLePaMo10,SqPi+13}. 
Newman et al.~\cite{NeFoBa02} show that the fraction of such edges in commonly studied graphs is quite high (refer also to \Tab{properties}),
and subsequent studies underlined the importance of such edges in network
formation and information diffusion~\cite{GaLo04, MiKoGuDrBh08, KwLePaMo10}. 
The set of wedges (\Fig{dwedge}) and triangles (\Fig{dtri}) involving directed and reciprocal edges  holds information about the underlying dynamics~\cite{HoLe70,Milo2002,Fa07,Fa10,SonKanKim12,SqGa12}.
But it is challenging to use this information to compare different graphs. 
We treat a directed graph as having two different types of edges: directed and reciprocal.
A reciprocal edge is technically a pair of directed edges, $\{(i,j),(j,i)\}$, that we treat as a single
reciprocal edge. In our figures, reciprocal edges are depicted as double-headed arrows. 
Treating reciprocal edges explicitly has been done since the \emph{triad census} work of Holland and Leinhardt~\cite{HoLe70},
and more recently for trade network analyses~\cite{GaLo06,SqGa12}.
Observe that reciprocal edges are essentially undirected.
In this paper, the \emph{total number of edges} refers to the sum of the number of directed edges plus the number of reciprocal edges.
Following \cite{NeFoBa02}, we define \emph{reciprocity} of a graph, $r$, as the ratio of  the number of reciprocal edges to the total
number of edges.  
\begin{figure}[t]
\centering
  \begin{tikzpicture}[nd/.style={circle,draw,fill=teal!50,inner sep=2pt},framed]    
    \matrix[column sep=0.4cm, row sep=0.2cm,ampersand replacement=\&]
    {
      \directedwedge{\linkba}{\linkba}{(i)}{\wout} \&
      \directedwedge{\linkab}{\linkba}{(ii)}{\wmid} \&
      \directedwedge{\linkab}{\linkab}{(iii)}{\win} \\
      \directedwedge{\linkaba}{\linkba}{(iv)}{\wrecipout} \&
      \directedwedge{\linkaba}{\linkab}{(v)}{\wrecipin} \&
      \directedwedge{\linkaba}{\linkaba}{(vi)}{\wreciptot} \\
    };
  \end{tikzpicture}  
\caption{Directed wedges}
\label{fig:dwedge}
\end{figure}
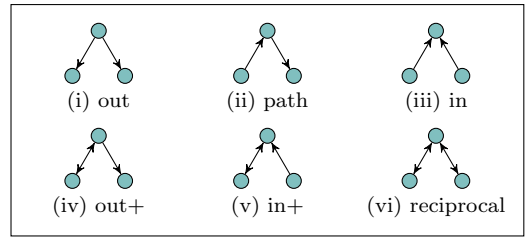
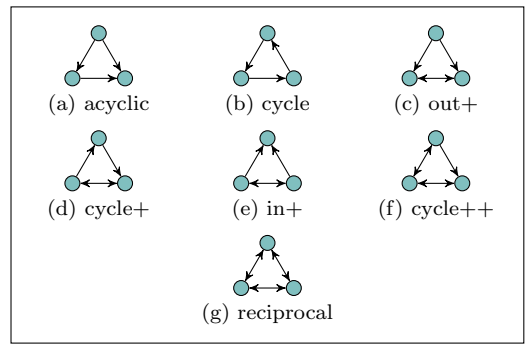
\begin{figure}[t]
\centering
  \begin{tikzpicture}[nd/.style={circle,draw,fill=teal!50,inner sep=2pt},framed]    
    \matrix[column sep=0.4cm, row sep=0.2cm,ampersand replacement=\&]
    {
      \directedtriangle{\linkab}{\linkba}{\linkba}{(a)}{\ttrans} \&
      \directedtriangle{\linkab}{\linkba}{\linkab}{(b)}{\tcycle} \&
      \directedtriangle{\linkaba}{\linkba}{\linkba}{(c)}{\toutrecip} \\
      \directedtriangle{\linkaba}{\linkab}{\linkba}{(d)}{\tmidrecip} \&
      \directedtriangle{\linkaba}{\linkab}{\linkab}{(e)}{\tinrecip} \&
      \directedtriangle{\linkaba}{\linkaba}{\linkba}{(f)}{\ttworecip} \\
      \& \directedtriangle{\linkaba}{\linkaba}{\linkaba}{(g)}{\tthreerecip} \&\\
    };
  \end{tikzpicture}
  \caption{Directed triangles}
    \label{fig:dtri}
\end{figure}
\subsection{Main results of this paper}
We generalize the classic notion
of \emph{transitivity} (pg. 243 of~\cite{WaFa94}), also called  the global clustering coefficient, to directed graphs. 
We say a wedge is \emph{closed} if it participates in a triangle. 
As described formally in \Sec{dir}, considering all possible directed wedge and triangle combinations yields a set of 15 directed closure values that provide a triadic summary of a directed graph. 
We perform experiments on a set of publicly available datasets and present the directed closure
information in a succinct form that allows for a comparison of different graphs. This leads to a series
of observations.
\begin{asparaitem}
	\item \textbf{Heterogeneity of closure:} We find the closure fractions of wedges vary greatly depending
	on the wedge type. ``In'' wedges are typically the most numerous but are rarely closed.
	On the other hand, \emph{all other} wedge types are much less numerous but are closed at a higher rate.
	\item \textbf{Reciprocity inducing closure:} For every graph we analyze, the presence of a reciprocal
	edge in a wedge greatly increases the probability of closure. In other words, wedges with reciprocal edges participate
	in triangles more frequently than wedges without any reciprocity. 
	\item \textbf{Transitivity correlates with reciprocity:}  Certain triangles are infrequent, specifically \ttrans and \tmidrecip  triangles. 
	The fact that cycles without reciprocal edges are so rare
	suggests that transitivity and reciprocity go hand in hand.
	This appears to validate the importance
	of transitivity, as posited by Holland and Leinhardt~\cite{HoLe70} in the social science community.
	\item \textbf{Directed closures not explained by degree structure:} It has been observed that many
	network patterns can be explained by the degree structure. Configuration models
	involving reciprocity were introduced by Garlaschelli and Loffredo~\cite{GaLo06} and also studied in \cite{DuKo+12}. These models have been
	observed to give better predictions of triadic structure~\cite{SqGa12,FaSqGa13}. However, we mathematically 
	and empirically verify that these models do not explain the directed closure patterns of real networks.
	\end{asparaitem}
\subsection{Previous work}
The earliest study of directed triads with reciprocity is in the social sciences,
by Holland and Leinhardt~\cite{HoLe70}. They compute the \emph{triad census} that counts the 16 different possible triads 
(including the 3 patterns with at most one edge). They also try to measure the effects of reciprocity 
in network formation.
Skvoretz et al.~\cite{Sk90,SkFaAg04}
use the triad census of predict various biases in network formation. 
In a more recent study, Faust~\cite{Fa10} compares the structure of various graphs using the triad census. 
Most of this work has been restricted to small
data sets having only a few hundred nodes. Counting triads has been referred to as \emph{motif finding}
in the bioinformatics community~\cite{Milo2002}. 
Directed triangle counts have been used
to define enhanced modularity measures~\cite{SeArGo11}.
Simpler versions of triad census counts have also
been used to analyze gaming data~\cite{SonKanKim12}. 
Szell et al.\@ also perform triadic analysis on gaming data~\cite{SzTh10,SzLaTh10}.
A classic \emph{local} measure of triangle density is the \emph{clustering coefficient}, introduced by Watts and Strogatz~\cite{WaSt98}.
Fagiolo~\cite{Fa07} proposes a local clustering coefficient measure for directed networks, though it ignores reciprocity.
Ahnert and Fink~\cite{AhFi08} construct ``clustering coefficients signatures" from these measures and classify directed networks. 
Recent work of Winkler and Reichardt~\cite{WiRe13} discusses the occurrence of the 16 different induced subgraphs on 3 vertices.
They give an ingeneous model based on Steiner Triple Systems that can match more nuances of triad counts than other models.
(But their model does not match the degree distributions.)
A configuration model explicitly modeling reciprocity was given by Garlaschelli and Loffredo~\cite{GaLo06} and also studied in \cite{DuKo+12}. There is further work showing its ability to match triadic patterns in trade networks~\cite{SqGa12,FaSqGa13}.
Contrary to this work, our results show that for massive networks (like web and social networks), these models are unable to match triadic closure patterns.
We have more discussion in \Sec{null}.
\section{Directed closures values} \label{sec:dir}
We begin with some notation and introduction to the directed structures in \Fig{dwedge} and \Fig{dtri}.
We stress that these types form a partition of all wedges and triangles. Since reciprocal
edges are distinguished, we do not think of (say) the \wrecipout wedge containing an \wout wedge. 
We use 
\begin{align*}
  \psi & \in \set{{\rm i},{\rm ii},\dots,{\rm vi}} = \text{wedge type, and}\\
  \tau & \in \set{{\rm a}, {\rm b}, \dots, {\rm g}} = \text{triangle type}.
\end{align*}
Different triangle types naturally contain
different types of wedges. This information is summarized by the function 
\begin{displaymath}
  \chi(\psi,\tau) = \text{number of $\psi$ wedges  in a $\tau$ triangle}.
\end{displaymath}
A table showing all values of $\chi(\psi,\tau)$ is provided
in \Tab{triangle_by_wedge}. 
There are 15 nonzero entries in this table, and zeros are omitted for clarity.
\begin{table}[thp] 
\centering 
\begin{tabular}{c|c|cccccc|} 
\multicolumn{1}{c}{}
 & \multicolumn{7}{c}{Wedge type ($\psi$)} \\
\cline{2-8} 
\multirow{7}{*}{\vspace{-1.1in}\hspace*{-0.1in}\begin{rotate}{90}Triangle type ($\tau$)\end{rotate}}
 &\;\;\;\;\;  &\; i\; &\; ii\; &\; iii \; &\;  iv \; & \; v \; & \; vi \; \\  \cline{2-8} 
 & a & 1 & 1 & 1 &   &   &   \\ \cline{2-8}
 & b &   & 3 &   &   &   &   \\ \cline{2-8}
 & c & 1 &   &   &   & 2 &   \\ \cline{2-8}
 & d &   & 1 &   & 1 & 1 &   \\ \cline{2-8}
 & e &   &   & 1 & 2 &   &   \\ \cline{2-8}
 & f &   &   &   & 1 & 1 & 1 \\ \cline{2-8}
 & g &   &   &   &   &   & 3 \\
\cline{2-8} % inserts single-line
\end{tabular}
  \caption{Wedges per triangle: $\chi(\psi,\tau)$.} 
\label{tab:triangle_by_wedge}
\end{table}
Given the degree of a node, we can calculate the number of each type of wedge it participates in.	
For vertex $v$, let 
\begin{displaymath}
  W_{v,\psi} = \set{ \text{ $\psi$ wedges centered at node $v$ }}
\end{displaymath}
We compute $|W_{v,\psi}|$ given the degrees of $v$. Let
\begin{align*}
  \din{v} & = \text{ indegree of $v$}, \\
  \dout{v} &= \text{ outdegree of $v$, and } \\
  \drec{v} &= \text{ reciprocal degree of $v$}.
\end{align*}
From these values, we can calculate $|W_{v,\psi}|$ for any vertex, as summarized in \Tab{wedge_counts}.
\setlength{\tabcolsep}{3pt}
\begin{table}[htp] 
\centering 
\begin{tabular}{c|cccccc} 
$\psi$ & \dwedge{\linkba}{\linkba}{$v$}  & \dwedge{\linkab}{\linkba}{$v$} &\dwedge{\linkab}{\linkab}{$v$}   &\dwedge{\linkaba}{\linkba}{$v$}  & \dwedge{\linkaba}{\linkab}{$v$} & \dwedge{\linkaba}{\linkaba}{$v$} \\\hline
$|W_{v,\psi}|$  
& $\displaystyle {\dout{v} \choose 2}$
& $\displaystyle \din{v}\dout{v}$ 
& $\displaystyle {\din{v} \choose 2}$   
& $\displaystyle \dout{v}\drec{v}$ 
& $\displaystyle \din{v}\drec{v}$ 
& $\displaystyle {\drec{v} \choose 2}$   \\
\end{tabular}
  \caption{Number of $\psi$ wedges per vertex}
\label{tab:wedge_counts}
\end{table}
\setlength{\tabcolsep}{0pt}
To define the directed closure values, we first define
\begin{align*}
  W_{\psi} & = \bigcup_{v} W_{v,\psi} = \text{set of all wedges of type $\psi$, and }\\
  T_{\tau} & = \text{set of all triangles of type $\tau$}.
\end{align*}
We define the directed $(\psi,\tau)$-closure, $\gcc_{\psi,\tau}$,
as the fraction of $\psi$-wedges that are $\tau$-closed, i.e.,
\begin{displaymath}
  \gcc_{\psi,\tau} = \frac{\chi(\psi,\tau)\; |T_\tau|}{|W_\psi|}.
\end{displaymath}
Note that if a type $\tau$ triangle contains no type $\psi$ wedge, then
this quantity is zero because of $\chi(\psi,\tau)$.
This definition is consistent with the undirected notion of
transitivity. If we let $W$ and $T$ denote the set of all wedges and
triangles, respectively, in an undirected graph, then the transitivity
is defined as $\kappa = 3|T|/|W|$. In this case, we know that three
wedges participate in every triangle, which is the analogue for $\chi(\psi,\tau)$.
\section{Observations on directed closure values}
We analyze the directed closure properties of various real graphs, whose properties are presented in \Tab{properties}. In this table,   
$r$ denotes the reciprocity and $\kappa$ denotes 
the undirected transitivity.
\subsection{Representations}
Figures~\ref{fig:web-Google}--\ref{fig:soc-Slashdot0902} illustrate the $\gcc_{\psi,\tau}$ values.
We explain using the example of web-Google~\cite{Snap} in \Fig{web-Google}.
The percentage of each triangle is shown at the top of each figure,
along with the color code for the triangle, e.g., 10\% of the
triangles in web-Google are reciprocal, and these are represented by
the color pink.
The percentage of each wedge type is shown along the $x$-axis. For
instance, 90\% of the wedges in web-Google are in wedges.
The height of the bar above each wedge denotes the rate at which that
wedge is closed. For instance, 50\% of the out wedges close for
web-Google.
The color codes on the bars show the type of triangle that the wedge
becomes. For instance, the vast majority of closed out+ wedges in
web-Google become in+ triangles (per the green color).
Finally, the transitivity ($\gcc$) of the undirected graph is marked
by a thick dashed line. For web-Google, $\gcc = 0.055$.
\setlength{\tabcolsep}{3pt}
\begin{table}[t!]
\caption{Properties of the graphs}
\label{tab:properties}
\centering
{\footnotesize
\begin{tabular}{rrrrr}
\multicolumn{1}{c}{
Graph Name} &	\multicolumn{1}{c}{\ Vertices\ }& 	\multicolumn{1}{c}{\ Edges\ } & \multicolumn{1}{c}{\ $r$\ } & \multicolumn{1}{c}{\ $\gcc$\ }   \\\hline 
amazon0505	& 410K&	3357K& 0.55	&	0.162	\\
soc-Slashdot0902 &	82K&	870K	& 0.84 &	0.024\\
web-Stanford	&282K&	2312K &	0.28 & 0.009\\
web-BerkStan&	685K	&7601K&	0.25 &	0.007\\
wiki-Talk	&2394K&	5021K& 0.14 & 0.002 \\
web-Google&	876K&	5105K&	0.31&	0.055\\
soc-Epinions1&	76K&	509K&	0.41	& 0.066 \\
web-NotreDame&	326K&	1470K&	0.52&	0.088 \\
youtube-links&	1158K &	4945K&	0.79&	0.006 \\
flickr-links	&1861K &	22614K&	0.62&	0.112\\
soc-livejournal&\ 	5284K\  &\ 76938K\ &\ 0.73\ &\  0.124\ \\ \hline 
\end{tabular}
}
\end{table} 
\begin{figure}
	\input{tikz-figures/web-Google-cc.tikz}
	\caption{Directed closure for web-Google}
\label{fig:web-Google}
\vspace{30pt}
	\input{tikz-figures/web-Stanford-cc.tikz}
	\caption{Directed closure for web-Stanford}
\label{fig:web-Stanford}
\vspace{30pt}
	\input{tikz-figures/web-BerkStan-cc.tikz}
	\caption{Directed closure for web-BerkStan}
\label{fig:web-BerkStan}
\label{fig:all-web}
\end{figure}
\begin{figure}
	\input{tikz-figures/soc-Epinions1-cc.tikz}
\caption{Directed closure for soc-Epinions1}
\label{fig:soc-Epinions1}
\vspace{30pt}
	\input{tikz-figures/livejournal-cc.tikz}
	\caption{Directed closure for livejournal}
\label{fig:livejournal}
\vspace{30pt}
	\input{tikz-figures/soc-Slashdot0902-cc.tikz}
	\caption{Directed closure for soc-Slashdot0902}
\label{fig:soc-Slashdot0902}
\end{figure}
\subsection{Similarities of directed closure rates} \label{sec:sim}
Figures \ref{fig:web-Google}--\ref{fig:web-BerkStan} show the closure charts for three different  web graphs:
web-Google, web-Stanford, and web-BerkStan~\cite{Snap}. These graphs
have vertices for web pages and directed edges for web links. Figures \ref{fig:soc-Epinions1}--\ref{fig:soc-Slashdot0902}
have the charts for three social networks~\cite{Snap}. The vertices of soc-Epinions correspond to the members
of Epinions, a consumer review site. A directed edge between users shows a trust relationship originating
from one user (these are signed by trust/distrust, which we ignore).
The vertices of soc-Slashdot~\cite{Snap} are users and edges represent tagging as friend or foe.
The vertices of soc-livejournal~\cite{ChKu+09,Law} are Slashdot users with edges denoting friendship (which is one-way).
Observe the similarity of the closure rates and proportions of the different wedges for the web graphs, despite them being from different
sources (and different sizes). The color patterns are remarkably similar, showing similar distributions
of different closures. The social networks show more variation, but the overall structure of the charts
is not far from the web graphs. In general, we note that in wedges rarely close and reciprocal and out+
wedges generally close the most frequently. 
\subsection{Heterogeneity of closure} \label{sec:het}
The heterogeneity of wedge closure is quite clear from all the closure charts. 
In the web graphs, the in wedges rarely close while all other wedges close at a rate of 25\% or higher.
On the other hand, the in wedges are the most frequent (90\% or more) so the
undirected transitivity is
always below 0.05.
The heterogenity is not as dramatic in the social networks, but there is some variation
in closures over the wedge types. Consistently, in wedges close at the lowest rate and \wreciptot wedges
close most frequently.
\subsection{Effect of reciprocity on closure rates}
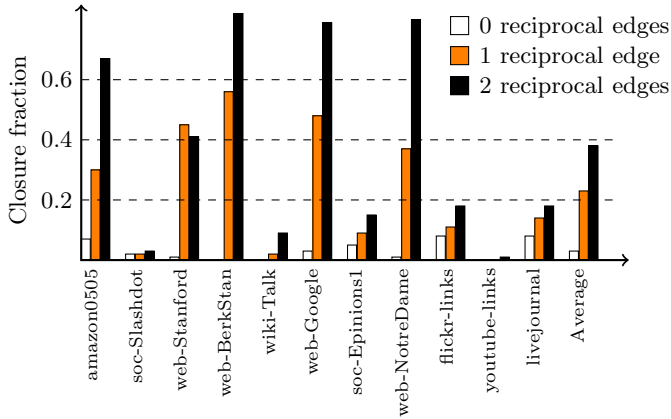
\begin{figure}[htbp]
\centering
\input{tikz-figures/recip.tikz}
\caption{Closure rates computed according to the number of reciprocal edges per wedge.}
\label{fig:reciprocity}
\end{figure}

We can see from Figures~\ref{fig:web-Google}--\ref{fig:soc-Slashdot0902} that wedges with reciprocal edges appear to close more frequently than those without. We do a comprehensive
calculation on a variety of graphs in \Fig{reciprocity} to show what proportion of wedges with $k \in \set{0,1,2}$ reciprocal
edges closes into any triangle. Observe the strong influence of reciprocation in the closure rates.
The average of chance of closure for a wedge without reciprocal edges  is only 3\%,   but this number goes to 23\%  if one of the edges is reciprocal and further increases to 38\% when both edges are reciprocal.  This finding is consistent with the earlier reports about  reciprocal edges, indicating stronger ties between two vertices~\cite{NeFoBa02,GaLo04, MiKoGuDrBh08, KwLePaMo10}. It also underscores how important it is to consider direction in networks since the rate of wedge closure depends on reciprocity of its edges.
\subsection{The connection between reciprocity and cycles}
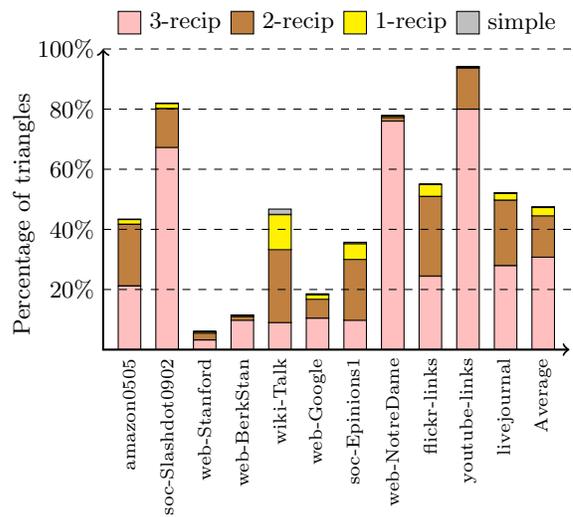
\begin{figure}[htbp]
\centering
 \input{tikz-figures/cycles.tikz}
\caption{Proportion of cycles in triangles, broken down by the number of reciprocal links.}
\label{fig:cycles}
\end{figure}

Throughout Figures~\ref{fig:web-Google}--\ref{fig:soc-Slashdot0902}, we notice the infrequency of \tcycle{} and \tmidrecip{} triangles (colored light blue and yellow, respectively), totaling less than 6\% in all cases. These are two of the four triangles that contain a cycle, the other two being \ttworecip{} and  \tthreerecip{} triangles (brown and pink, respectively). 
It is common to assume that a cycle indicates a strong
tie between three vertices, and so we might hypothesize that reciprocation is expected. This is exactly what we see in \Fig{cycles}, where we illustrate the proportion of triangles that are cycles and the breakdown among the types. Almost all triangles with a cycle are either \ttworecip{} or \tthreerecip{} triangles.
We almost never see any \tcycle{} triangles. Again, this is more evidence that reciprocal edges
play an important role in graph structure.
The results demonstrate  the power of transitivity of real world networks: social relationships carried forward two steps (as a transitive
relation) almost always lead to reciprocation.
\section{Directed closures are not explained by degrees} \label{sec:null}
\begin{figure}[htbp]
\centering
 \input{tikz-figures/true-vs-pred.tikz}
\caption{True vs.\@ predicted ratio for several directed closures. (For convenience, we truncate at a factor of 1000.)}
\label{fig:true-pred}
\end{figure}
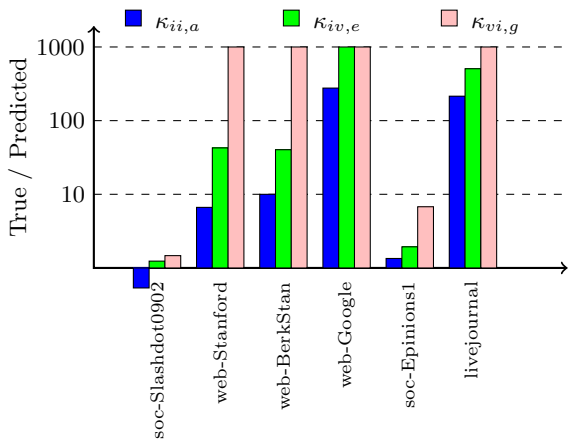

\begin{figure}[h]
\centering
 \vspace{10pt}
 \input{tikz-figures/z-score.tikz}
\caption{Z-scores for several directed closures. (For convenience, we truncate all values at a factor of 1000.)}
\label{fig:z-score}
\end{figure}
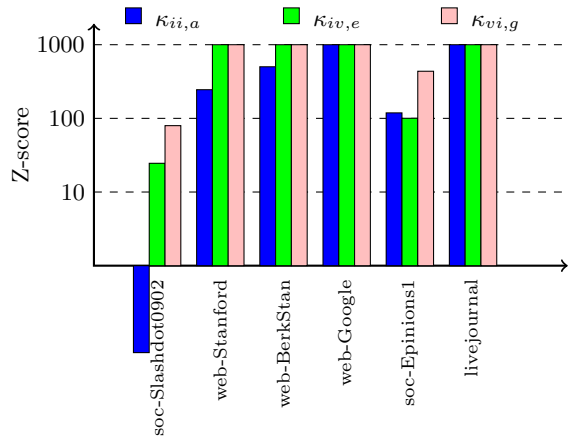

Could the different directed closures rates simply be a consequence of the (directed and reciprocal) degrees?
We assert that this is \emph{not} the case, by showing that a configuration model
accounting for reciprocity cannot generate the high $(\psi,\tau)$-closure values seen in the previous section.
We explain the reciprocal configuration model of Garlaschelli and
Loffredo~\cite{GaLo06} which respects both directed and reciprocal
degrees.
We assume that $\din{v}$, $\dout{v}$, and $\drec{v}$ are specified for every node.
Let $\mins = \sum_v \din{v}$ or $\mout = \sum_v \dout{v}$ denote the number
of directed edges, and let $\mrec = \sum_v \drec{v}$ be twice the number of reciprocal edges.
Conceptually, we assume that the model 
has an edge $i \rightarrow j$ with probability $\dout{i}\din{j}/\mout$,
and by a reciprocal edge between $i$ and $j$ with probability $\drec{i}\drec{j}/\mrec$.
(We stress this is only an approximation, and there can be significant deviations from this
approximation, as pointed out previously~\cite{NePa03}.)
We give an approximation for expected $\gcc_{\psi,\tau}$, values. Previous analyses for transitivities on configuration graphs
(see Newman~\cite{Ne03}, Park and Newman~\cite{NePa03}) focus on undirected graphs, and it is not hard to generalize these calculations.
Focus on the proportion of path wedges that become acyclic triangles, i.e., $\gcc_{ii,a}$, since it is consistently large for all networks we experiment on.
Note that $\chi(ii,a) = 1$, so $\gcc_{ii,a} = |T_{a}|/|W_{ii}|$.
Observe that $\EX[|W_{ii}|] = \sum_v \din{v}\dout{v}$; furthermore,
$|W_{ii}|$ is well-concentrated, so it is almost always close to its mean~\cite{bdm2006}.
By linearity of expectation $\EX[|T_{d}|] = \sum_{i,j,k} \dout{i}\dout{i}\dout{j}\din{j}\din{k}\din{k}/(\mout)^3$
(also explicitly given by Squartini and Garlaschelli~\cite{SqGa12}).
From this, we surmise
\begin{eqnarray*}
\EX[\gcc_{ii,a}] & = & \frac{\sum_{i,j,k} \dout{i}\dout{i}\dout{j}\din{j}\din{k}\din{k}}{(\mout)^3\sum_v \din{v}\dout{v}} \\
& = & \frac{(\sum_j \dout{j}\din{j}) (\sum_{i,k} \dout{i}\dout{i}\din{k}\din{k})}{(\mout)^3 (\sum_v \din{v}\dout{v})} \\
& = & \frac{\sum_i \dout{i}\dout{i} \sum_k \din{k}\din{k}}{(\mout)^3} \\
& \approx & \frac{4|W_{(i)}|\cdot|W_{(iii)}|}{(\mout)^3}.
\end{eqnarray*}
We also look at the proportion of out+ wedges that become in triangles, i.e., $\EX[\gcc_{iv,e}]$, another closure that is quite large. We have $|W_{iv}| \sum_v \dout{v}\drec{v}$.
We note that $2|T_{e}| \approx \sum_{i,j,k} \dout{i}\drec{i}\dout{j}\drec{j} (\din{k})^2/(\mout)^2\mrec$.
(We get the factor of two because type-e triangles have two type-iv wedges). Hence,
\begin{eqnarray*}
\EX[\gcc_{iv,e}] & = & \frac{\sum_{i,j,k} \dout{i}\drec{i}\dout{j}\drec{j}\din{k}\din{k}}{(\mout)^2\mrec\sum_v \dout{v}\drec{v}} \\
& = & \frac{\sum_{j} \dout{j}\drec{j} \sum_k \din{k}\din{k}}{(\mout)^2\mrec} \\
& \approx & \frac{|W_{iv}|\cdot 2|W_{(iii)}|}{(\mout)^2\mrec}.
\end{eqnarray*}
Finally, consider reciprocal wedges that become reciprocal triangles, i.e., $\EX[\gcc_{vi,g}]$, another closure that is often large.
This is equivalent to an undirected calculation involving only reciprocal edges.
Hence
\begin{eqnarray*}
\EX[\gcc_{vi,g}] \approx \frac{4|W_{vi}|^2}{(\mrec)^3}.
\end{eqnarray*}
In \Fig{true-pred}, we show the ratio of the predicted closure values given by the equations above and
the true closure values for the six networks we have been analyzing. Other
than the graph soc-Slashdot0902, all other predictions are much smaller than the true value. For the web networks,
the prediction is many order of magnitudes smaller than the true value. This is strong evidence
that the directed closures cannot be explained by the degrees alone.
To show the statistical significance of these results, we compute the $Z$-scores for the results of the configuration model as compared to the true data.
We observe the value of a directed closure value for multiple realizations of the configuration model. Let $X$ denotes that random variable with mean $\mu$ and variance $\sigma$. The
Z-score is $(X-\mu)/\sigma$, i.e., the number of standard deviations the observation is from the mean of the models. We empirically estimate the mean and variance 
 by generating 100 random instances from the configuration
model. We present the results in \Fig{z-score}. Other
than for $\gcc_{ii,a}$ in soc-Slashdot0902, \emph{all other $Z$-scores are positive}. 
This is in excellent agreement with \Fig{true-pred} and
our mathematical argument that the configuration model underestimates the directed
closures (except in one case). Observe that the $Z$-scores are extremely large, which is an
indication of statistical significance of the directed closure values for real networks.
\section{Conclusions}
We perform a detailed study of directed triangles in massive networks, by defining the set of directed closure measures. These quantities reveal
a surprising amount of information about directed graphs. We observe heterogeneity in closure rates of different wedges, the impact of reciprocity on closure rates, 
and the power of transitivity in the structure  of triangles. 
We mathematically and empirically justify the statistical significance of these measures on real networks.
We argue that is of great interest to design network models that recreate the directed closure patterns of real-world networks.
\begin{acknowledgments}
This work was funded by the GRAPHS Program at DARPA, the applied mathematics program at the United
States Department of Energy,
and by an Early Career Award from the Laboratory
Directed Research \& Development (LDRD) program at Sandia National
Laboratories. 
Sandia National Laboratories is a multi-program laboratory managed and operated by Sandia Corporation, a wholly owned subsidiary of Lockheed Martin Corporation, for the U.S. Department of Energy's National Nuclear Security Administration under contract DE-AC04-94AL85000.
\end{acknowledgments}
\bibliographystyle{apsrev4-1}
\bibliography{directedCC}  
\end{document}

%% file: tikz-figures/web-Google-cc.tikz
\begin{tikzpicture} [scale=1.400000e+000]

\draw [thick] [<->] (0,2.3) -- (0,0) -- (3.8,0);
\draw (-0.1,0.5) -- (0.1,0.5);
\node [left] at (-0.1,0.5) {0.2};
\draw (-0.1,1) -- (0.1,1);
\node [left] at (-0.1,1) {0.4};
\draw (-0.1,1.5) -- (0.1,1.5);
\node [left] at (-0.1,1.5) {0.6};
\draw (-0.1,2) -- (0.1,2);
\node [left] at (-0.1,2) {0.8};
\node [rotate=90, scale=1] at (-0.700000,1.150000) {Directed Closure};
\node (A) [scale=0.6] at (0.5,-0.2) [vertex] {};
\node (B) [scale=0.6] at (0.35,-0.5) [vertex] {};
\node (C) [scale=0.6] at (0.65,-0.5) [vertex] {};
\draw [-stealth'] (A) -- (B);
\draw [-stealth'] (A) -- (C);
\node [scale=1] at (0.5,-0.7) {2\%};
\node (A) [scale=0.6] at (1.1,-0.2) [vertex] {};
\node (B) [scale=0.6] at (0.95,-0.5) [vertex] {};
\node (C) [scale=0.6] at (1.25,-0.5) [vertex] {};
\draw [stealth'-] (A) -- (B);
\draw [-stealth'] (A) -- (C);
\node [scale=1] at (1.1,-0.7) {2\%};
\node (A) [scale=0.6] at (1.7,-0.2) [vertex] {};
\node (B) [scale=0.6] at (1.55,-0.5) [vertex] {};
\node (C) [scale=0.6] at (1.85,-0.5) [vertex] {};
\draw [stealth'-] (A) -- (B);
\draw [stealth'-] (A) -- (C);
\node [scale=1] at (1.7,-0.7) {90\%};
\node (A) [scale=0.6] at (2.3,-0.2) [vertex] {};
\node (B) [scale=0.6] at (2.15,-0.5) [vertex] {};
\node (C) [scale=0.6] at (2.45,-0.5) [vertex] {};
\draw [stealth'-stealth'] (A) -- (B);
\draw [-stealth'] (A) -- (C);
\node [scale=1] at (2.3,-0.7) {1\%};
\node (A) [scale=0.6] at (2.9,-0.2) [vertex] {};
\node (B) [scale=0.6] at (2.75,-0.5) [vertex] {};
\node (C) [scale=0.6] at (3.05,-0.5) [vertex] {};
\draw [stealth'-stealth'] (A) -- (B);
\draw [stealth'-] (A) -- (C);
\node [scale=1] at (2.9,-0.7) {3\%};
\node (A) [scale=0.6] at (3.5,-0.2) [vertex] {};
\node (B) [scale=0.6] at (3.35,-0.5) [vertex] {};
\node (C) [scale=0.6] at (3.65,-0.5) [vertex] {};
\draw [stealth'-stealth'] (A) -- (B);
\draw [stealth'-stealth'] (A) -- (C);
\node [scale=1] at (3.5,-0.7) {1\%};
\draw [fill=\typea,thin] (0.3,0) rectangle (0.7,0.668731);
\draw [fill=\typec,thin] (0.3,0.668731) rectangle (0.7,1.25457);
\draw [fill=\typea,thin] (0.9,0) rectangle (1.3,0.732648);
\draw [fill=\typeb,thin] (0.9,0.732648) rectangle (1.3,0.748475);
\draw [fill=\typed,thin] (0.9,0.748475) rectangle (1.3,0.779002);
\draw [fill=\typea,thin] (1.5,0) rectangle (1.9,0.0181914);
\draw [fill=\typee,thin] (1.5,0.0181914) rectangle (1.9,0.0255706);
\draw [fill=\typed,thin] (2.1,0) rectangle (2.5,0.0702175);
\draw [fill=\typee,thin] (2.1,0.0702175) rectangle (2.5,1.43743);
\draw [fill=\typef,thin] (2.1,1.43743) rectangle (2.5,1.73707);
\draw [fill=\typec,thin] (2.7,0) rectangle (3.1,0.913801);
\draw [fill=\typed,thin] (2.7,0.913801) rectangle (3.1,0.935531);
\draw [fill=\typef,thin] (2.7,0.935531) rectangle (3.1,1.02826);
\draw [fill=\typef,thin] (3.3,0) rectangle (3.7,0.331632);
\draw [fill=\typeg,thin] (3.3,0.331632) rectangle (3.7,1.97441);
\draw [fill=\typea,thin] (-0.15,2.6) rectangle (0.45,2.85);
\node (A) [scale=0.75] at (0.15,3.4) [vertex] {};
\node (B) [scale=0.75] at (-0.05,3) [vertex] {};
\node (C) [scale=0.75] at (0.35,3) [vertex] {};
\draw [-stealth'] (A) -- (B);
\draw [-stealth'] (A) -- (C);
\draw [-stealth'] (B) -- (C);
\node [scale=1] at (0.15,2.45) {36\%};
\draw [fill=\typeb,thin] (0.45,2.6) rectangle (1.05,2.85);
\node (A) [scale=0.75] at (0.75,3.4) [vertex] {};
\node (B) [scale=0.75] at (0.55,3) [vertex] {};
\node (C) [scale=0.75] at (0.95,3) [vertex] {};
\draw [-stealth'] (A) -- (B);
\draw [stealth'-] (A) -- (C);
\draw [-stealth'] (B) -- (C);
\node [scale=1] at (0.75,2.45) {$< 1$\%};
\draw [fill=\typec,thin] (1.05,2.6) rectangle (1.65,2.85);
\node (A) [scale=0.75] at (1.35,3.4) [vertex] {};
\node (B) [scale=0.75] at (1.15,3) [vertex] {};
\node (C) [scale=0.75] at (1.55,3) [vertex] {};
\draw [-stealth'] (A) -- (B);
\draw [-stealth'] (A) -- (C);
\draw [stealth'-stealth'] (B) -- (C);
\node [scale=1] at (1.35,2.45) {31\%};
\draw [fill=\typed,thin] (1.65,2.6) rectangle (2.25,2.85);
\node (A) [scale=0.75] at (1.95,3.4) [vertex] {};
\node (B) [scale=0.75] at (1.75,3) [vertex] {};
\node (C) [scale=0.75] at (2.15,3) [vertex] {};
\draw [stealth'-] (A) -- (B);
\draw [-stealth'] (A) -- (C);
\draw [stealth'-stealth'] (B) -- (C);
\node [scale=1] at (1.95,2.45) {1\%};
\draw [fill=\typee,thin] (2.25,2.6) rectangle (2.85,2.85);
\node (A) [scale=0.75] at (2.55,3.4) [vertex] {};
\node (B) [scale=0.75] at (2.35,3) [vertex] {};
\node (C) [scale=0.75] at (2.75,3) [vertex] {};
\draw [stealth'-] (A) -- (B);
\draw [stealth'-] (A) -- (C);
\draw [stealth'-stealth'] (B) -- (C);
\node [scale=1] at (2.55,2.45) {14\%};
\draw [fill=\typef,thin] (2.85,2.6) rectangle (3.45,2.85);
\node (A) [scale=0.75] at (3.15,3.4) [vertex] {};
\node (B) [scale=0.75] at (2.95,3) [vertex] {};
\node (C) [scale=0.75] at (3.35,3) [vertex] {};
\draw [stealth'-stealth'] (A) -- (B);
\draw [-stealth'] (A) -- (C);
\draw [stealth'-stealth'] (B) -- (C);
\node [scale=1] at (3.15,2.45) {6\%};
\draw [fill=\typeg,thin] (3.45,2.6) rectangle (4.05,2.85);
\node (A) [scale=0.75] at (3.75,3.4) [vertex] {};
\node (B) [scale=0.75] at (3.55,3) [vertex] {};
\node (C) [scale=0.75] at (3.95,3) [vertex] {};
\draw [stealth'-stealth'] (A) -- (B);
\draw [stealth'-stealth'] (A) -- (C);
\draw [stealth'-stealth'] (B) -- (C);
\node [scale=1] at (3.75,2.45) {10\%};
\draw [dashed, very thick] (-0.1,0.138077) -- (3.8,0.138077);
\node [left] at (-0.1,0.138077) {$\kappa$};
\end{tikzpicture}

%% file: tikz-figures/web-Stanford-cc.tikz
\begin{tikzpicture} [scale=1.400000e+000]

\draw [thick] [<->] (0,2.3) -- (0,0) -- (3.8,0);
\draw (-0.1,0.5) -- (0.1,0.5);
\node [left] at (-0.1,0.5) {0.2};
\draw (-0.1,1) -- (0.1,1);
\node [left] at (-0.1,1) {0.4};
\draw (-0.1,1.5) -- (0.1,1.5);
\node [left] at (-0.1,1.5) {0.6};
\draw (-0.1,2) -- (0.1,2);
\node [left] at (-0.1,2) {0.8};
\node [rotate=90, scale=1] at (-0.700000,1.150000) {Directed Closure};
\node (A) [scale=0.6] at (0.5,-0.2) [vertex] {};
\node (B) [scale=0.6] at (0.35,-0.5) [vertex] {};
\node (C) [scale=0.6] at (0.65,-0.5) [vertex] {};
\draw [-stealth'] (A) -- (B);
\draw [-stealth'] (A) -- (C);
\node [scale=1] at (0.5,-0.7) {$< 1$\%};
\node (A) [scale=0.6] at (1.1,-0.2) [vertex] {};
\node (B) [scale=0.6] at (0.95,-0.5) [vertex] {};
\node (C) [scale=0.6] at (1.25,-0.5) [vertex] {};
\draw [stealth'-] (A) -- (B);
\draw [-stealth'] (A) -- (C);
\node [scale=1] at (1.1,-0.7) {$< 1$\%};
\node (A) [scale=0.6] at (1.7,-0.2) [vertex] {};
\node (B) [scale=0.6] at (1.55,-0.5) [vertex] {};
\node (C) [scale=0.6] at (1.85,-0.5) [vertex] {};
\draw [stealth'-] (A) -- (B);
\draw [stealth'-] (A) -- (C);
\node [scale=1] at (1.7,-0.7) {98\%};
\node (A) [scale=0.6] at (2.3,-0.2) [vertex] {};
\node (B) [scale=0.6] at (2.15,-0.5) [vertex] {};
\node (C) [scale=0.6] at (2.45,-0.5) [vertex] {};
\draw [stealth'-stealth'] (A) -- (B);
\draw [-stealth'] (A) -- (C);
\node [scale=1] at (2.3,-0.7) {$< 1$\%};
\node (A) [scale=0.6] at (2.9,-0.2) [vertex] {};
\node (B) [scale=0.6] at (2.75,-0.5) [vertex] {};
\node (C) [scale=0.6] at (3.05,-0.5) [vertex] {};
\draw [stealth'-stealth'] (A) -- (B);
\draw [stealth'-] (A) -- (C);
\node [scale=1] at (2.9,-0.7) {1\%};
\node (A) [scale=0.6] at (3.5,-0.2) [vertex] {};
\node (B) [scale=0.6] at (3.35,-0.5) [vertex] {};
\node (C) [scale=0.6] at (3.65,-0.5) [vertex] {};
\draw [stealth'-stealth'] (A) -- (B);
\draw [stealth'-stealth'] (A) -- (C);
\node [scale=1] at (3.5,-0.7) {$< 1$\%};
\draw [fill=\typea,thin] (0.3,0) rectangle (0.7,0.723838);
\draw [fill=\typec,thin] (0.3,0.723838) rectangle (0.7,1.2137);
\draw [fill=\typea,thin] (0.9,0) rectangle (1.3,1.06753);
\draw [fill=\typeb,thin] (0.9,1.06753) rectangle (1.3,1.07268);
\draw [fill=\typed,thin] (0.9,1.07268) rectangle (1.3,1.08283);
\draw [fill=\typea,thin] (1.5,0) rectangle (1.9,0.00360676);
\draw [fill=\typee,thin] (1.5,0.00360676) rectangle (1.9,0.00440671);
\draw [fill=\typed,thin] (2.1,0) rectangle (2.5,0.0363383);
\draw [fill=\typee,thin] (2.1,0.0363383) rectangle (2.5,1.73203);
\draw [fill=\typef,thin] (2.1,1.73203) rectangle (2.5,1.91002);
\draw [fill=\typec,thin] (2.7,0) rectangle (3.1,0.948294);
\draw [fill=\typed,thin] (2.7,0.948294) rectangle (3.1,0.954954);
\draw [fill=\typef,thin] (2.7,0.954954) rectangle (3.1,0.987575);
\draw [fill=\typef,thin] (3.3,0) rectangle (3.7,0.19521);
\draw [fill=\typeg,thin] (3.3,0.19521) rectangle (3.7,1.01709);
\draw [fill=\typea,thin] (-0.15,2.6) rectangle (0.45,2.85);
\node (A) [scale=0.75] at (0.15,3.4) [vertex] {};
\node (B) [scale=0.75] at (-0.05,3) [vertex] {};
\node (C) [scale=0.75] at (0.35,3) [vertex] {};
\draw [-stealth'] (A) -- (B);
\draw [-stealth'] (A) -- (C);
\draw [-stealth'] (B) -- (C);
\node [scale=1] at (0.15,2.45) {49\%};
\draw [fill=\typeb,thin] (0.45,2.6) rectangle (1.05,2.85);
\node (A) [scale=0.75] at (0.75,3.4) [vertex] {};
\node (B) [scale=0.75] at (0.55,3) [vertex] {};
\node (C) [scale=0.75] at (0.95,3) [vertex] {};
\draw [-stealth'] (A) -- (B);
\draw [stealth'-] (A) -- (C);
\draw [-stealth'] (B) -- (C);
\node [scale=1] at (0.75,2.45) {$< 1$\%};
\draw [fill=\typec,thin] (1.05,2.6) rectangle (1.65,2.85);
\node (A) [scale=0.75] at (1.35,3.4) [vertex] {};
\node (B) [scale=0.75] at (1.15,3) [vertex] {};
\node (C) [scale=0.75] at (1.55,3) [vertex] {};
\draw [-stealth'] (A) -- (B);
\draw [-stealth'] (A) -- (C);
\draw [stealth'-stealth'] (B) -- (C);
\node [scale=1] at (1.35,2.45) {33\%};
\draw [fill=\typed,thin] (1.65,2.6) rectangle (2.25,2.85);
\node (A) [scale=0.75] at (1.95,3.4) [vertex] {};
\node (B) [scale=0.75] at (1.75,3) [vertex] {};
\node (C) [scale=0.75] at (2.15,3) [vertex] {};
\draw [stealth'-] (A) -- (B);
\draw [-stealth'] (A) -- (C);
\draw [stealth'-stealth'] (B) -- (C);
\node [scale=1] at (1.95,2.45) {$< 1$\%};
\draw [fill=\typee,thin] (2.25,2.6) rectangle (2.85,2.85);
\node (A) [scale=0.75] at (2.55,3.4) [vertex] {};
\node (B) [scale=0.75] at (2.35,3) [vertex] {};
\node (C) [scale=0.75] at (2.75,3) [vertex] {};
\draw [stealth'-] (A) -- (B);
\draw [stealth'-] (A) -- (C);
\draw [stealth'-stealth'] (B) -- (C);
\node [scale=1] at (2.55,2.45) {11\%};
\draw [fill=\typef,thin] (2.85,2.6) rectangle (3.45,2.85);
\node (A) [scale=0.75] at (3.15,3.4) [vertex] {};
\node (B) [scale=0.75] at (2.95,3) [vertex] {};
\node (C) [scale=0.75] at (3.35,3) [vertex] {};
\draw [stealth'-stealth'] (A) -- (B);
\draw [-stealth'] (A) -- (C);
\draw [stealth'-stealth'] (B) -- (C);
\node [scale=1] at (3.15,2.45) {2\%};
\draw [fill=\typeg,thin] (3.45,2.6) rectangle (4.05,2.85);
\node (A) [scale=0.75] at (3.75,3.4) [vertex] {};
\node (B) [scale=0.75] at (3.55,3) [vertex] {};
\node (C) [scale=0.75] at (3.95,3) [vertex] {};
\draw [stealth'-stealth'] (A) -- (B);
\draw [stealth'-stealth'] (A) -- (C);
\draw [stealth'-stealth'] (B) -- (C);
\node [scale=1] at (3.75,2.45) {3\%};
\draw [dashed, very thick] (-0.1,0.021544) -- (3.8,0.021544);
\node [left] at (-0.1,0.021544) {$\kappa$};
\end{tikzpicture}

%% file: tikz-figures/web-BerkStan-cc.tikz
\begin{tikzpicture} [scale=1.400000e+000]

\draw [thick] [<->] (0,2.3) -- (0,0) -- (3.8,0);
\draw (-0.1,0.5) -- (0.1,0.5);
\node [left] at (-0.1,0.5) {0.2};
\draw (-0.1,1) -- (0.1,1);
\node [left] at (-0.1,1) {0.4};
\draw (-0.1,1.5) -- (0.1,1.5);
\node [left] at (-0.1,1.5) {0.6};
\draw (-0.1,2) -- (0.1,2);
\node [left] at (-0.1,2) {0.8};
\node [rotate=90, scale=1] at (-0.700000,1.150000) {Directed Closure};
\node (A) [scale=0.6] at (0.5,-0.2) [vertex] {};
\node (B) [scale=0.6] at (0.35,-0.5) [vertex] {};
\node (C) [scale=0.6] at (0.65,-0.5) [vertex] {};
\draw [-stealth'] (A) -- (B);
\draw [-stealth'] (A) -- (C);
\node [scale=1] at (0.5,-0.7) {$< 1$\%};
\node (A) [scale=0.6] at (1.1,-0.2) [vertex] {};
\node (B) [scale=0.6] at (0.95,-0.5) [vertex] {};
\node (C) [scale=0.6] at (1.25,-0.5) [vertex] {};
\draw [stealth'-] (A) -- (B);
\draw [-stealth'] (A) -- (C);
\node [scale=1] at (1.1,-0.7) {$< 1$\%};
\node (A) [scale=0.6] at (1.7,-0.2) [vertex] {};
\node (B) [scale=0.6] at (1.55,-0.5) [vertex] {};
\node (C) [scale=0.6] at (1.85,-0.5) [vertex] {};
\draw [stealth'-] (A) -- (B);
\draw [stealth'-] (A) -- (C);
\node [scale=1] at (1.7,-0.7) {99\%};
\node (A) [scale=0.6] at (2.3,-0.2) [vertex] {};
\node (B) [scale=0.6] at (2.15,-0.5) [vertex] {};
\node (C) [scale=0.6] at (2.45,-0.5) [vertex] {};
\draw [stealth'-stealth'] (A) -- (B);
\draw [-stealth'] (A) -- (C);
\node [scale=1] at (2.3,-0.7) {$< 1$\%};
\node (A) [scale=0.6] at (2.9,-0.2) [vertex] {};
\node (B) [scale=0.6] at (2.75,-0.5) [vertex] {};
\node (C) [scale=0.6] at (3.05,-0.5) [vertex] {};
\draw [stealth'-stealth'] (A) -- (B);
\draw [stealth'-] (A) -- (C);
\node [scale=1] at (2.9,-0.7) {$< 1$\%};
\node (A) [scale=0.6] at (3.5,-0.2) [vertex] {};
\node (B) [scale=0.6] at (3.35,-0.5) [vertex] {};
\node (C) [scale=0.6] at (3.65,-0.5) [vertex] {};
\draw [stealth'-stealth'] (A) -- (B);
\draw [stealth'-stealth'] (A) -- (C);
\node [scale=1] at (3.5,-0.7) {$< 1$\%};
\draw [fill=\typea,thin] (0.3,0) rectangle (0.7,0.733668);
\draw [fill=\typec,thin] (0.3,0.733668) rectangle (0.7,1.43663);
\draw [fill=\typea,thin] (0.9,0) rectangle (1.3,1.22449);
\draw [fill=\typeb,thin] (0.9,1.22449) rectangle (1.3,1.22802);
\draw [fill=\typed,thin] (0.9,1.22802) rectangle (1.3,1.24114);
\draw [fill=\typea,thin] (1.5,0) rectangle (1.9,0.00216544);
\draw [fill=\typee,thin] (1.5,0.00216544) rectangle (1.9,0.00309248);
\draw [fill=\typed,thin] (2.1,0) rectangle (2.5,0.0269656);
\draw [fill=\typee,thin] (2.1,0.0269656) rectangle (2.5,2.18138);
\draw [fill=\typef,thin] (2.1,2.18138) rectangle (2.5,2.26179);
\draw [fill=\typec,thin] (2.7,0) rectangle (3.1,1.16161);
\draw [fill=\typed,thin] (2.7,1.16161) rectangle (3.1,1.1681);
\draw [fill=\typef,thin] (2.7,1.1681) rectangle (3.1,1.18747);
\draw [fill=\typef,thin] (3.3,0) rectangle (3.7,0.0785902);
\draw [fill=\typeg,thin] (3.3,0.0785902) rectangle (3.7,2.04014);
\draw [fill=\typea,thin] (-0.15,2.6) rectangle (0.45,2.85);
\node (A) [scale=0.75] at (0.15,3.4) [vertex] {};
\node (B) [scale=0.75] at (-0.05,3) [vertex] {};
\node (C) [scale=0.75] at (0.35,3) [vertex] {};
\draw [-stealth'] (A) -- (B);
\draw [-stealth'] (A) -- (C);
\draw [-stealth'] (B) -- (C);
\node [scale=1] at (0.15,2.45) {37\%};
\draw [fill=\typeb,thin] (0.45,2.6) rectangle (1.05,2.85);
\node (A) [scale=0.75] at (0.75,3.4) [vertex] {};
\node (B) [scale=0.75] at (0.55,3) [vertex] {};
\node (C) [scale=0.75] at (0.95,3) [vertex] {};
\draw [-stealth'] (A) -- (B);
\draw [stealth'-] (A) -- (C);
\draw [-stealth'] (B) -- (C);
\node [scale=1] at (0.75,2.45) {$< 1$\%};
\draw [fill=\typec,thin] (1.05,2.6) rectangle (1.65,2.85);
\node (A) [scale=0.75] at (1.35,3.4) [vertex] {};
\node (B) [scale=0.75] at (1.15,3) [vertex] {};
\node (C) [scale=0.75] at (1.55,3) [vertex] {};
\draw [-stealth'] (A) -- (B);
\draw [-stealth'] (A) -- (C);
\draw [stealth'-stealth'] (B) -- (C);
\node [scale=1] at (1.35,2.45) {36\%};
\draw [fill=\typed,thin] (1.65,2.6) rectangle (2.25,2.85);
\node (A) [scale=0.75] at (1.95,3.4) [vertex] {};
\node (B) [scale=0.75] at (1.75,3) [vertex] {};
\node (C) [scale=0.75] at (2.15,3) [vertex] {};
\draw [stealth'-] (A) -- (B);
\draw [-stealth'] (A) -- (C);
\draw [stealth'-stealth'] (B) -- (C);
\node [scale=1] at (1.95,2.45) {$< 1$\%};
\draw [fill=\typee,thin] (2.25,2.6) rectangle (2.85,2.85);
\node (A) [scale=0.75] at (2.55,3.4) [vertex] {};
\node (B) [scale=0.75] at (2.35,3) [vertex] {};
\node (C) [scale=0.75] at (2.75,3) [vertex] {};
\draw [stealth'-] (A) -- (B);
\draw [stealth'-] (A) -- (C);
\draw [stealth'-stealth'] (B) -- (C);
\node [scale=1] at (2.55,2.45) {16\%};
\draw [fill=\typef,thin] (2.85,2.6) rectangle (3.45,2.85);
\node (A) [scale=0.75] at (3.15,3.4) [vertex] {};
\node (B) [scale=0.75] at (2.95,3) [vertex] {};
\node (C) [scale=0.75] at (3.35,3) [vertex] {};
\draw [stealth'-stealth'] (A) -- (B);
\draw [-stealth'] (A) -- (C);
\draw [stealth'-stealth'] (B) -- (C);
\node [scale=1] at (3.15,2.45) {1\%};
\draw [fill=\typeg,thin] (3.45,2.6) rectangle (4.05,2.85);
\node (A) [scale=0.75] at (3.75,3.4) [vertex] {};
\node (B) [scale=0.75] at (3.55,3) [vertex] {};
\node (C) [scale=0.75] at (3.95,3) [vertex] {};
\draw [stealth'-stealth'] (A) -- (B);
\draw [stealth'-stealth'] (A) -- (C);
\draw [stealth'-stealth'] (B) -- (C);
\node [scale=1] at (3.75,2.45) {10\%};
\draw [dashed, very thick] (-0.1,0.0173385) -- (3.8,0.0173385);
\node [left] at (-0.1,0.0173385) {$\kappa$};
\end{tikzpicture}

%% file: tikz-figures/soc-Epinions1-cc.tikz
\begin{tikzpicture} [scale=1.400000e+000]

\draw [thick] [<->] (0,2.3) -- (0,0) -- (3.8,0);
\draw (-0.1,0.5) -- (0.1,0.5);
\node [left] at (-0.1,0.5) {0.05};
\draw (-0.1,1) -- (0.1,1);
\node [left] at (-0.1,1) {0.1};
\draw (-0.1,1.5) -- (0.1,1.5);
\node [left] at (-0.1,1.5) {0.15};
\draw (-0.1,2) -- (0.1,2);
\node [left] at (-0.1,2) {0.2};
\node [rotate=90, scale=1] at (-0.800000,1.150000) {Directed Closure};
\node (A) [scale=0.6] at (0.5,-0.2) [vertex] {};
\node (B) [scale=0.6] at (0.35,-0.5) [vertex] {};
\node (C) [scale=0.6] at (0.65,-0.5) [vertex] {};
\draw [-stealth'] (A) -- (B);
\draw [-stealth'] (A) -- (C);
\node [scale=1] at (0.5,-0.7) {17\%};
\node (A) [scale=0.6] at (1.1,-0.2) [vertex] {};
\node (B) [scale=0.6] at (0.95,-0.5) [vertex] {};
\node (C) [scale=0.6] at (1.25,-0.5) [vertex] {};
\draw [stealth'-] (A) -- (B);
\draw [-stealth'] (A) -- (C);
\node [scale=1] at (1.1,-0.7) {11\%};
\node (A) [scale=0.6] at (1.7,-0.2) [vertex] {};
\node (B) [scale=0.6] at (1.55,-0.5) [vertex] {};
\node (C) [scale=0.6] at (1.85,-0.5) [vertex] {};
\draw [stealth'-] (A) -- (B);
\draw [stealth'-] (A) -- (C);
\node [scale=1] at (1.7,-0.7) {37\%};
\node (A) [scale=0.6] at (2.3,-0.2) [vertex] {};
\node (B) [scale=0.6] at (2.15,-0.5) [vertex] {};
\node (C) [scale=0.6] at (2.45,-0.5) [vertex] {};
\draw [stealth'-stealth'] (A) -- (B);
\draw [-stealth'] (A) -- (C);
\node [scale=1] at (2.3,-0.7) {12\%};
\node (A) [scale=0.6] at (2.9,-0.2) [vertex] {};
\node (B) [scale=0.6] at (2.75,-0.5) [vertex] {};
\node (C) [scale=0.6] at (3.05,-0.5) [vertex] {};
\draw [stealth'-stealth'] (A) -- (B);
\draw [stealth'-] (A) -- (C);
\node [scale=1] at (2.9,-0.7) {16\%};
\node (A) [scale=0.6] at (3.5,-0.2) [vertex] {};
\node (B) [scale=0.6] at (3.35,-0.5) [vertex] {};
\node (C) [scale=0.6] at (3.65,-0.5) [vertex] {};
\draw [stealth'-stealth'] (A) -- (B);
\draw [stealth'-stealth'] (A) -- (C);
\node [scale=1] at (3.5,-0.7) {7\%};
\draw [fill=\typea,thin] (0.3,0) rectangle (0.7,0.420325);
\draw [fill=\typec,thin] (0.3,0.420325) rectangle (0.7,0.642695);
\draw [fill=\typea,thin] (0.9,0) rectangle (1.3,0.66796);
\draw [fill=\typeb,thin] (0.9,0.66796) rectangle (1.3,0.696835);
\draw [fill=\typed,thin] (0.9,0.696835) rectangle (1.3,0.802919);
\draw [fill=\typea,thin] (1.5,0) rectangle (1.9,0.195209);
\draw [fill=\typee,thin] (1.5,0.195209) rectangle (1.9,0.280391);
\draw [fill=\typed,thin] (2.1,0) rectangle (2.5,0.0925817);
\draw [fill=\typee,thin] (2.1,0.0925817) rectangle (2.5,0.601329);
\draw [fill=\typef,thin] (2.1,0.601329) rectangle (2.5,0.961277);
\draw [fill=\typec,thin] (2.7,0) rectangle (3.1,0.472532);
\draw [fill=\typed,thin] (2.7,0.472532) rectangle (3.1,0.543459);
\draw [fill=\typef,thin] (2.7,0.543459) rectangle (3.1,0.819216);
\draw [fill=\typef,thin] (3.3,0) rectangle (3.7,0.610283);
\draw [fill=\typeg,thin] (3.3,0.610283) rectangle (3.7,1.50371);
\draw [fill=\typea,thin] (-0.15,2.6) rectangle (0.45,2.85);
\node (A) [scale=0.75] at (0.15,3.4) [vertex] {};
\node (B) [scale=0.75] at (-0.05,3) [vertex] {};
\node (C) [scale=0.75] at (0.35,3) [vertex] {};
\draw [-stealth'] (A) -- (B);
\draw [-stealth'] (A) -- (C);
\draw [-stealth'] (B) -- (C);
\node [scale=1] at (0.15,2.45) {33\%};
\draw [fill=\typeb,thin] (0.45,2.6) rectangle (1.05,2.85);
\node (A) [scale=0.75] at (0.75,3.4) [vertex] {};
\node (B) [scale=0.75] at (0.55,3) [vertex] {};
\node (C) [scale=0.75] at (0.95,3) [vertex] {};
\draw [-stealth'] (A) -- (B);
\draw [stealth'-] (A) -- (C);
\draw [-stealth'] (B) -- (C);
\node [scale=1] at (0.75,2.45) {$< 1$\%};
\draw [fill=\typec,thin] (1.05,2.6) rectangle (1.65,2.85);
\node (A) [scale=0.75] at (1.35,3.4) [vertex] {};
\node (B) [scale=0.75] at (1.15,3) [vertex] {};
\node (C) [scale=0.75] at (1.55,3) [vertex] {};
\draw [-stealth'] (A) -- (B);
\draw [-stealth'] (A) -- (C);
\draw [stealth'-stealth'] (B) -- (C);
\node [scale=1] at (1.35,2.45) {17\%};
\draw [fill=\typed,thin] (1.65,2.6) rectangle (2.25,2.85);
\node (A) [scale=0.75] at (1.95,3.4) [vertex] {};
\node (B) [scale=0.75] at (1.75,3) [vertex] {};
\node (C) [scale=0.75] at (2.15,3) [vertex] {};
\draw [stealth'-] (A) -- (B);
\draw [-stealth'] (A) -- (C);
\draw [stealth'-stealth'] (B) -- (C);
\node [scale=1] at (1.95,2.45) {5\%};
\draw [fill=\typee,thin] (2.25,2.6) rectangle (2.85,2.85);
\node (A) [scale=0.75] at (2.55,3.4) [vertex] {};
\node (B) [scale=0.75] at (2.35,3) [vertex] {};
\node (C) [scale=0.75] at (2.75,3) [vertex] {};
\draw [stealth'-] (A) -- (B);
\draw [stealth'-] (A) -- (C);
\draw [stealth'-stealth'] (B) -- (C);
\node [scale=1] at (2.55,2.45) {14\%};
\draw [fill=\typef,thin] (2.85,2.6) rectangle (3.45,2.85);
\node (A) [scale=0.75] at (3.15,3.4) [vertex] {};
\node (B) [scale=0.75] at (2.95,3) [vertex] {};
\node (C) [scale=0.75] at (3.35,3) [vertex] {};
\draw [stealth'-stealth'] (A) -- (B);
\draw [-stealth'] (A) -- (C);
\draw [stealth'-stealth'] (B) -- (C);
\node [scale=1] at (3.15,2.45) {20\%};
\draw [fill=\typeg,thin] (3.45,2.6) rectangle (4.05,2.85);
\node (A) [scale=0.75] at (3.75,3.4) [vertex] {};
\node (B) [scale=0.75] at (3.55,3) [vertex] {};
\node (C) [scale=0.75] at (3.95,3) [vertex] {};
\draw [stealth'-stealth'] (A) -- (B);
\draw [stealth'-stealth'] (A) -- (C);
\draw [stealth'-stealth'] (B) -- (C);
\node [scale=1] at (3.75,2.45) {10\%};
\draw [dashed, very thick] (-0.1,0.656789) -- (3.8,0.656789);
\node [left] at (-0.1,0.656789) {$\kappa$};
\end{tikzpicture}

%% file: tikz-figures/livejournal-cc.tikz
\begin{tikzpicture} [scale=1.400000e+000]

\draw [thick] [<->] (0,2.3) -- (0,0) -- (3.8,0);
\draw (-0.1,0.5) -- (0.1,0.5);
\node [left] at (-0.1,0.5) {0.05};
\draw (-0.1,1) -- (0.1,1);
\node [left] at (-0.1,1) {0.1};
\draw (-0.1,1.5) -- (0.1,1.5);
\node [left] at (-0.1,1.5) {0.15};
\draw (-0.1,2) -- (0.1,2);
\node [left] at (-0.1,2) {0.2};
\node [rotate=90, scale=1] at (-0.800000,1.150000) {Directed Closure};
\node (A) [scale=0.6] at (0.5,-0.2) [vertex] {};
\node (B) [scale=0.6] at (0.35,-0.5) [vertex] {};
\node (C) [scale=0.6] at (0.65,-0.5) [vertex] {};
\draw [-stealth'] (A) -- (B);
\draw [-stealth'] (A) -- (C);
\node [scale=1] at (0.5,-0.7) {8\%};
\node (A) [scale=0.6] at (1.1,-0.2) [vertex] {};
\node (B) [scale=0.6] at (0.95,-0.5) [vertex] {};
\node (C) [scale=0.6] at (1.25,-0.5) [vertex] {};
\draw [stealth'-] (A) -- (B);
\draw [-stealth'] (A) -- (C);
\node [scale=1] at (1.1,-0.7) {5\%};
\node (A) [scale=0.6] at (1.7,-0.2) [vertex] {};
\node (B) [scale=0.6] at (1.55,-0.5) [vertex] {};
\node (C) [scale=0.6] at (1.85,-0.5) [vertex] {};
\draw [stealth'-] (A) -- (B);
\draw [stealth'-] (A) -- (C);
\node [scale=1] at (1.7,-0.7) {32\%};
\node (A) [scale=0.6] at (2.3,-0.2) [vertex] {};
\node (B) [scale=0.6] at (2.15,-0.5) [vertex] {};
\node (C) [scale=0.6] at (2.45,-0.5) [vertex] {};
\draw [stealth'-stealth'] (A) -- (B);
\draw [-stealth'] (A) -- (C);
\node [scale=1] at (2.3,-0.7) {13\%};
\node (A) [scale=0.6] at (2.9,-0.2) [vertex] {};
\node (B) [scale=0.6] at (2.75,-0.5) [vertex] {};
\node (C) [scale=0.6] at (3.05,-0.5) [vertex] {};
\draw [stealth'-stealth'] (A) -- (B);
\draw [stealth'-] (A) -- (C);
\node [scale=1] at (2.9,-0.7) {18\%};
\node (A) [scale=0.6] at (3.5,-0.2) [vertex] {};
\node (B) [scale=0.6] at (3.35,-0.5) [vertex] {};
\node (C) [scale=0.6] at (3.65,-0.5) [vertex] {};
\draw [stealth'-stealth'] (A) -- (B);
\draw [stealth'-stealth'] (A) -- (C);
\node [scale=1] at (3.5,-0.7) {24\%};
\draw [fill=\typea,thin] (0.3,0) rectangle (0.7,0.949808);
\draw [fill=\typec,thin] (0.3,0.949808) rectangle (0.7,1.63831);
\draw [fill=\typea,thin] (0.9,0) rectangle (1.3,1.48895);
\draw [fill=\typeb,thin] (0.9,1.48895) rectangle (1.3,1.50307);
\draw [fill=\typed,thin] (0.9,1.50307) rectangle (1.3,1.68497);
\draw [fill=\typea,thin] (1.5,0) rectangle (1.9,0.251981);
\draw [fill=\typee,thin] (1.5,0.251981) rectangle (1.9,0.445034);
\draw [fill=\typed,thin] (2.1,0) rectangle (2.5,0.0767675);
\draw [fill=\typee,thin] (2.1,0.0767675) rectangle (2.5,1.03959);
\draw [fill=\typef,thin] (2.1,1.03959) rectangle (2.5,1.75279);
\draw [fill=\typec,thin] (2.7,0) rectangle (3.1,0.637248);
\draw [fill=\typed,thin] (2.7,0.637248) rectangle (3.1,0.690949);
\draw [fill=\typef,thin] (2.7,0.690949) rectangle (3.1,1.18984);
\draw [fill=\typef,thin] (3.3,0) rectangle (3.7,0.373996);
\draw [fill=\typeg,thin] (3.3,0.373996) rectangle (3.7,1.81328);
\draw [fill=\typea,thin] (-0.15,2.6) rectangle (0.45,2.85);
\node (A) [scale=0.75] at (0.15,3.4) [vertex] {};
\node (B) [scale=0.75] at (-0.05,3) [vertex] {};
\node (C) [scale=0.75] at (0.35,3) [vertex] {};
\draw [-stealth'] (A) -- (B);
\draw [-stealth'] (A) -- (C);
\draw [-stealth'] (B) -- (C);
\node [scale=1] at (0.15,2.45) {19\%};
\draw [fill=\typeb,thin] (0.45,2.6) rectangle (1.05,2.85);
\node (A) [scale=0.75] at (0.75,3.4) [vertex] {};
\node (B) [scale=0.75] at (0.55,3) [vertex] {};
\node (C) [scale=0.75] at (0.95,3) [vertex] {};
\draw [-stealth'] (A) -- (B);
\draw [stealth'-] (A) -- (C);
\draw [-stealth'] (B) -- (C);
\node [scale=1] at (0.75,2.45) {$< 1$\%};
\draw [fill=\typec,thin] (1.05,2.6) rectangle (1.65,2.85);
\node (A) [scale=0.75] at (1.35,3.4) [vertex] {};
\node (B) [scale=0.75] at (1.15,3) [vertex] {};
\node (C) [scale=0.75] at (1.55,3) [vertex] {};
\draw [-stealth'] (A) -- (B);
\draw [-stealth'] (A) -- (C);
\draw [stealth'-stealth'] (B) -- (C);
\node [scale=1] at (1.35,2.45) {14\%};
\draw [fill=\typed,thin] (1.65,2.6) rectangle (2.25,2.85);
\node (A) [scale=0.75] at (1.95,3.4) [vertex] {};
\node (B) [scale=0.75] at (1.75,3) [vertex] {};
\node (C) [scale=0.75] at (2.15,3) [vertex] {};
\draw [stealth'-] (A) -- (B);
\draw [-stealth'] (A) -- (C);
\draw [stealth'-stealth'] (B) -- (C);
\node [scale=1] at (1.95,2.45) {2\%};
\draw [fill=\typee,thin] (2.25,2.6) rectangle (2.85,2.85);
\node (A) [scale=0.75] at (2.55,3.4) [vertex] {};
\node (B) [scale=0.75] at (2.35,3) [vertex] {};
\node (C) [scale=0.75] at (2.75,3) [vertex] {};
\draw [stealth'-] (A) -- (B);
\draw [stealth'-] (A) -- (C);
\draw [stealth'-stealth'] (B) -- (C);
\node [scale=1] at (2.55,2.45) {15\%};
\draw [fill=\typef,thin] (2.85,2.6) rectangle (3.45,2.85);
\node (A) [scale=0.75] at (3.15,3.4) [vertex] {};
\node (B) [scale=0.75] at (2.95,3) [vertex] {};
\node (C) [scale=0.75] at (3.35,3) [vertex] {};
\draw [stealth'-stealth'] (A) -- (B);
\draw [-stealth'] (A) -- (C);
\draw [stealth'-stealth'] (B) -- (C);
\node [scale=1] at (3.15,2.45) {22\%};
\draw [fill=\typeg,thin] (3.45,2.6) rectangle (4.05,2.85);
\node (A) [scale=0.75] at (3.75,3.4) [vertex] {};
\node (B) [scale=0.75] at (3.55,3) [vertex] {};
\node (C) [scale=0.75] at (3.95,3) [vertex] {};
\draw [stealth'-stealth'] (A) -- (B);
\draw [stealth'-stealth'] (A) -- (C);
\draw [stealth'-stealth'] (B) -- (C);
\node [scale=1] at (3.75,2.45) {28\%};
\draw [dashed, very thick] (-0.1,1.24041) -- (3.8,1.24041);
\node [left] at (-0.1,1.24041) {$\kappa$};
\end{tikzpicture}

%% file: tikz-figures/soc-Slashdot0902-cc.tikz
\begin{tikzpicture} [scale=1.400000e+000]

\draw [thick] [<->] (0,2.3) -- (0,0) -- (3.8,0);
\draw (-0.1,0.5) -- (0.1,0.5);
\node [left] at (-0.1,0.5) {0.01};
\draw (-0.1,1) -- (0.1,1);
\node [left] at (-0.1,1) {0.02};
\draw (-0.1,1.5) -- (0.1,1.5);
\node [left] at (-0.1,1.5) {0.03};
\draw (-0.1,2) -- (0.1,2);
\node [left] at (-0.1,2) {0.04};
\node [rotate=90, scale=1] at (-0.800000,1.150000) {Directed Closure};
\node (A) [scale=0.6] at (0.5,-0.2) [vertex] {};
\node (B) [scale=0.6] at (0.35,-0.5) [vertex] {};
\node (C) [scale=0.6] at (0.65,-0.5) [vertex] {};
\draw [-stealth'] (A) -- (B);
\draw [-stealth'] (A) -- (C);
\node [scale=1] at (0.5,-0.7) {6\%};
\node (A) [scale=0.6] at (1.1,-0.2) [vertex] {};
\node (B) [scale=0.6] at (0.95,-0.5) [vertex] {};
\node (C) [scale=0.6] at (1.25,-0.5) [vertex] {};
\draw [stealth'-] (A) -- (B);
\draw [-stealth'] (A) -- (C);
\node [scale=1] at (1.1,-0.7) {1\%};
\node (A) [scale=0.6] at (1.7,-0.2) [vertex] {};
\node (B) [scale=0.6] at (1.55,-0.5) [vertex] {};
\node (C) [scale=0.6] at (1.85,-0.5) [vertex] {};
\draw [stealth'-] (A) -- (B);
\draw [stealth'-] (A) -- (C);
\node [scale=1] at (1.7,-0.7) {5\%};
\node (A) [scale=0.6] at (2.3,-0.2) [vertex] {};
\node (B) [scale=0.6] at (2.15,-0.5) [vertex] {};
\node (C) [scale=0.6] at (2.45,-0.5) [vertex] {};
\draw [stealth'-stealth'] (A) -- (B);
\draw [-stealth'] (A) -- (C);
\node [scale=1] at (2.3,-0.7) {10\%};
\node (A) [scale=0.6] at (2.9,-0.2) [vertex] {};
\node (B) [scale=0.6] at (2.75,-0.5) [vertex] {};
\node (C) [scale=0.6] at (3.05,-0.5) [vertex] {};
\draw [stealth'-stealth'] (A) -- (B);
\draw [stealth'-] (A) -- (C);
\node [scale=1] at (2.9,-0.7) {13\%};
\node (A) [scale=0.6] at (3.5,-0.2) [vertex] {};
\node (B) [scale=0.6] at (3.35,-0.5) [vertex] {};
\node (C) [scale=0.6] at (3.65,-0.5) [vertex] {};
\draw [stealth'-stealth'] (A) -- (B);
\draw [stealth'-stealth'] (A) -- (C);
\node [scale=1] at (3.5,-0.7) {64\%};
\draw [fill=\typea,thin] (0.3,0) rectangle (0.7,0.150976);
\draw [fill=\typec,thin] (0.3,0.150976) rectangle (0.7,0.989337);
\draw [fill=\typea,thin] (0.9,0) rectangle (1.3,0.694902);
\draw [fill=\typeb,thin] (0.9,0.694902) rectangle (1.3,0.708991);
\draw [fill=\typed,thin] (0.9,0.708991) rectangle (1.3,1.23329);
\draw [fill=\typea,thin] (1.5,0) rectangle (1.9,0.178604);
\draw [fill=\typee,thin] (1.5,0.178604) rectangle (1.9,0.439773);
\draw [fill=\typed,thin] (2.1,0) rectangle (2.5,0.0668243);
\draw [fill=\typee,thin] (2.1,0.0668243) rectangle (2.5,0.325846);
\draw [fill=\typef,thin] (2.1,0.325846) rectangle (2.5,0.828736);
\draw [fill=\typec,thin] (2.7,0) rectangle (3.1,0.751076);
\draw [fill=\typed,thin] (2.7,0.751076) rectangle (3.1,0.802101);
\draw [fill=\typef,thin] (2.7,0.802101) rectangle (3.1,1.1861);
\draw [fill=\typef,thin] (3.3,0) rectangle (3.7,0.0806235);
\draw [fill=\typeg,thin] (3.3,0.0806235) rectangle (3.7,1.35068);
\draw [fill=\typea,thin] (-0.15,2.6) rectangle (0.45,2.85);
\node (A) [scale=0.75] at (0.15,3.4) [vertex] {};
\node (B) [scale=0.75] at (-0.05,3) [vertex] {};
\node (C) [scale=0.75] at (0.35,3) [vertex] {};
\draw [-stealth'] (A) -- (B);
\draw [-stealth'] (A) -- (C);
\draw [-stealth'] (B) -- (C);
\node [scale=1] at (0.15,2.45) {2\%};
\draw [fill=\typeb,thin] (0.45,2.6) rectangle (1.05,2.85);
\node (A) [scale=0.75] at (0.75,3.4) [vertex] {};
\node (B) [scale=0.75] at (0.55,3) [vertex] {};
\node (C) [scale=0.75] at (0.95,3) [vertex] {};
\draw [-stealth'] (A) -- (B);
\draw [stealth'-] (A) -- (C);
\draw [-stealth'] (B) -- (C);
\node [scale=1] at (0.75,2.45) {$< 1$\%};
\draw [fill=\typec,thin] (1.05,2.6) rectangle (1.65,2.85);
\node (A) [scale=0.75] at (1.35,3.4) [vertex] {};
\node (B) [scale=0.75] at (1.15,3) [vertex] {};
\node (C) [scale=0.75] at (1.55,3) [vertex] {};
\draw [-stealth'] (A) -- (B);
\draw [-stealth'] (A) -- (C);
\draw [stealth'-stealth'] (B) -- (C);
\node [scale=1] at (1.35,2.45) {13\%};
\draw [fill=\typed,thin] (1.65,2.6) rectangle (2.25,2.85);
\node (A) [scale=0.75] at (1.95,3.4) [vertex] {};
\node (B) [scale=0.75] at (1.75,3) [vertex] {};
\node (C) [scale=0.75] at (2.15,3) [vertex] {};
\draw [stealth'-] (A) -- (B);
\draw [-stealth'] (A) -- (C);
\draw [stealth'-stealth'] (B) -- (C);
\node [scale=1] at (1.95,2.45) {2\%};
\draw [fill=\typee,thin] (2.25,2.6) rectangle (2.85,2.85);
\node (A) [scale=0.75] at (2.55,3.4) [vertex] {};
\node (B) [scale=0.75] at (2.35,3) [vertex] {};
\node (C) [scale=0.75] at (2.75,3) [vertex] {};
\draw [stealth'-] (A) -- (B);
\draw [stealth'-] (A) -- (C);
\draw [stealth'-stealth'] (B) -- (C);
\node [scale=1] at (2.55,2.45) {3\%};
\draw [fill=\typef,thin] (2.85,2.6) rectangle (3.45,2.85);
\node (A) [scale=0.75] at (3.15,3.4) [vertex] {};
\node (B) [scale=0.75] at (2.95,3) [vertex] {};
\node (C) [scale=0.75] at (3.35,3) [vertex] {};
\draw [stealth'-stealth'] (A) -- (B);
\draw [-stealth'] (A) -- (C);
\draw [stealth'-stealth'] (B) -- (C);
\node [scale=1] at (3.15,2.45) {13\%};
\draw [fill=\typeg,thin] (3.45,2.6) rectangle (4.05,2.85);
\node (A) [scale=0.75] at (3.75,3.4) [vertex] {};
\node (B) [scale=0.75] at (3.55,3) [vertex] {};
\node (C) [scale=0.75] at (3.95,3) [vertex] {};
\draw [stealth'-stealth'] (A) -- (B);
\draw [stealth'-stealth'] (A) -- (C);
\draw [stealth'-stealth'] (B) -- (C);
\node [scale=1] at (3.75,2.45) {67\%};
\draw [dashed, very thick] (-0.1,1.20555) -- (3.8,1.20555);
\node [left] at (-0.1,1.20555) {$\kappa$};
\end{tikzpicture}

%% file: tikz-figures/recip.tikz
\begin{tikzpicture}\draw [fill=white,thin] (0.000,0) rectangle (0.120,0.280) ;
\draw [fill=orange,thin] (0.130,0) node[below]{\rotatebox[origin=t]{90}{\scriptsize amazon0505}} rectangle (0.250,1.200) ;
\draw [fill=black,thin] (0.260,0) rectangle (0.380,2.680) ;
\draw [fill=white,thin] (0.590,0) rectangle (0.710,0.080) ;
\draw [fill=orange,thin] (0.720,0) node[below]{\rotatebox[origin=t]{90}{\scriptsize soc-Slashdot}} rectangle (0.840,0.080) ;
\draw [fill=black,thin] (0.850,0) rectangle (0.970,0.120) ;
\draw [fill=white,thin] (1.180,0) rectangle (1.300,0.040) ;
\draw [fill=orange,thin] (1.310,0) node[below]{\rotatebox[origin=t]{90}{\scriptsize web-Stanford}} rectangle (1.430,1.800) ;
\draw [fill=black,thin] (1.440,0) rectangle (1.560,1.640) ;
\draw [fill=white,thin] (1.770,0) rectangle (1.890,0.000) ;
\draw [fill=orange,thin] (1.900,0) node[below]{\rotatebox[origin=t]{90}{\scriptsize web-BerkStan}} rectangle (2.020,2.240) ;
\draw [fill=black,thin] (2.030,0) rectangle (2.150,3.280) ;
\draw [fill=white,thin] (2.360,0) rectangle (2.480,0.000) ;
\draw [fill=orange,thin] (2.490,0) node[below]{\rotatebox[origin=t]{90}{\scriptsize wiki-Talk}} rectangle (2.610,0.080) ;
\draw [fill=black,thin] (2.620,0) rectangle (2.740,0.360) ;
\draw [fill=white,thin] (2.950,0) rectangle (3.070,0.120) ;
\draw [fill=orange,thin] (3.080,0) node[below]{\rotatebox[origin=t]{90}{\scriptsize web-Google}} rectangle (3.200,1.920) ;
\draw [fill=black,thin] (3.210,0) rectangle (3.330,3.160) ;
\draw [fill=white,thin] (3.540,0) rectangle (3.660,0.200) ;
\draw [fill=orange,thin] (3.670,0) node[below]{\rotatebox[origin=t]{90}{\scriptsize soc-Epinions1}} rectangle (3.790,0.360) ;
\draw [fill=black,thin] (3.800,0) rectangle (3.920,0.600) ;
\draw [fill=white,thin] (4.130,0) rectangle (4.250,0.040) ;
\draw [fill=orange,thin] (4.260,0) node[below]{\rotatebox[origin=t]{90}{\scriptsize web-NotreDame}} rectangle (4.380,1.480) ;
\draw [fill=black,thin] (4.390,0) rectangle (4.510,3.200) ;
\draw [fill=white,thin] (4.720,0) rectangle (4.840,0.320) ;
\draw [fill=orange,thin] (4.850,0) node[below]{\rotatebox[origin=t]{90}{\scriptsize flickr-links}} rectangle (4.970,0.440) ;
\draw [fill=black,thin] (4.980,0) rectangle (5.100,0.720) ;
\draw [fill=white,thin] (5.310,0) rectangle (5.430,0.000) ;
\draw [fill=orange,thin] (5.440,0) node[below]{\rotatebox[origin=t]{90}{\scriptsize youtube-links}} rectangle (5.560,0.000) ;
\draw [fill=black,thin] (5.570,0) rectangle (5.690,0.040) ;
\draw [fill=white,thin] (5.900,0) rectangle (6.020,0.320) ;
\draw [fill=orange,thin] (6.030,0) node[below]{\rotatebox[origin=t]{90}{\scriptsize livejournal}} rectangle (6.150,0.560) ;
\draw [fill=black,thin] (6.160,0) rectangle (6.280,0.720) ;
\draw [fill=white,thin] (6.490,0) rectangle (6.610,0.120) ;
\draw [fill=orange,thin] (6.620,0) node[below]{\rotatebox[origin=t]{90}{\scriptsize Average}} rectangle (6.740,0.920) ;
\draw [fill=black,thin] (6.750,0) rectangle (6.870,1.520) ;
\draw [<->, thick] (7.28,0) -- (0,0)-- (0,1.60) node[left]{\vspace{10ex}\hspace{-12ex}\rotatebox{90}{Closure fraction}} -- (0, 3.40) ;
\draw [dashed] (0, 0.800) node [left]{$0.2$} -- (7.100,0.800);
\draw [dashed] (0, 1.600) node [left]{$0.4$} -- (7.100,1.600);
\draw [dashed] (0, 2.400) node [left]{$0.6$} -- (4.510,2.400);
\draw [fill=white,thin] (4.91,3.00) node[right] at (4.8,3.1) {\hspace{2ex} \small 0 reciprocal edges} rectangle (5.15,3.24)  ;
\draw [fill=orange,thin] (4.91,2.60) node[right] at (4.8,2.7) {\hspace{2ex} \small 1 reciprocal edge} rectangle (5.15,2.84)  ;
\draw [fill=black,thin] (4.91,2.20) node[right] at (4.8,2.3)  {\hspace{2ex} \small 2 reciprocal edges} rectangle (5.15,2.44)  ;
\end{tikzpicture}

%% file: tikz-figures/cycles.tikz
\begin{tikzpicture}
\draw [fill=\typeg,thin] (0.200,0) node[below] at (0.35,0){\rotatebox[origin=t]{90}{\scriptsize amazon0505}} rectangle (0.500,0.847) ;
\draw [fill=\typef,thin] (0.200,0.85) rectangle (0.500,1.666) ;
\draw [fill=\typed,thin] (0.200,1.67) rectangle (0.500,1.734) ;
\draw [fill=\typeb,thin] (0.200,1.73) rectangle (0.500,1.735) ;
\draw [fill=\typeg,thin] (0.700,0) node[below] at (0.85,0){\rotatebox[origin=t]{90}{\scriptsize soc-Slashdot0902}} rectangle (1.000,2.694) ;
\draw [fill=\typef,thin] (0.700,2.69) rectangle (1.000,3.207) ;
\draw [fill=\typed,thin] (0.700,3.21) rectangle (1.000,3.275) ;
\draw [fill=\typeb,thin] (0.700,3.28) rectangle (1.000,3.276) ;
\draw [fill=\typeg,thin] (1.200,0) node[below] at (1.35,0){\rotatebox[origin=t]{90}{\scriptsize web-Stanford}} rectangle (1.500,0.129) ;
\draw [fill=\typef,thin] (1.200,0.13) rectangle (1.500,0.221) ;
\draw [fill=\typed,thin] (1.200,0.22) rectangle (1.500,0.240) ;
\draw [fill=\typeb,thin] (1.200,0.24) rectangle (1.500,0.243) ;
\draw [fill=\typeg,thin] (1.700,0) node[below] at (1.85,0){\rotatebox[origin=t]{90}{\scriptsize web-BerkStan}} rectangle (2.000,0.394) ;
\draw [fill=\typef,thin] (1.700,0.39) rectangle (2.000,0.442) ;
\draw [fill=\typed,thin] (1.700,0.44) rectangle (2.000,0.458) ;
\draw [fill=\typeb,thin] (1.700,0.46) rectangle (2.000,0.459) ;
\draw [fill=\typeg,thin] (2.200,0) node[below] at (2.35,0){\rotatebox[origin=t]{90}{\scriptsize wiki-Talk}} rectangle (2.500,0.364) ;
\draw [fill=\typef,thin] (2.200,0.36) rectangle (2.500,1.335) ;
\draw [fill=\typed,thin] (2.200,1.33) rectangle (2.500,1.795) ;
\draw [fill=\typeb,thin] (2.200,1.80) rectangle (2.500,1.870) ;
\draw [fill=\typeg,thin] (2.700,0) node[below] at (2.85,0){\rotatebox[origin=t]{90}{\scriptsize web-Google}} rectangle (3.000,0.419) ;
\draw [fill=\typef,thin] (2.700,0.42) rectangle (3.000,0.673) ;
\draw [fill=\typed,thin] (2.700,0.67) rectangle (3.000,0.732) ;
\draw [fill=\typeb,thin] (2.700,0.73) rectangle (3.000,0.743) ;
\draw [fill=\typeg,thin] (3.200,0) node[below] at (3.35,0){\rotatebox[origin=t]{90}{\scriptsize soc-Epinions1}} rectangle (3.500,0.394) ;
\draw [fill=\typef,thin] (3.200,0.39) rectangle (3.500,1.202) ;
\draw [fill=\typed,thin] (3.200,1.20) rectangle (3.500,1.410) ;
\draw [fill=\typeb,thin] (3.200,1.41) rectangle (3.500,1.429) ;
\draw [fill=\typeg,thin] (3.700,0) node[below] at (3.85,0){\rotatebox[origin=t]{90}{\scriptsize web-NotreDame}} rectangle (4.000,3.044) ;
\draw [fill=\typef,thin] (3.700,3.04) rectangle (4.000,3.092) ;
\draw [fill=\typed,thin] (3.700,3.09) rectangle (4.000,3.110) ;
\draw [fill=\typeb,thin] (3.700,3.11) rectangle (4.000,3.114) ;
\draw [fill=\typeg,thin] (4.200,0) node[below] at (4.35,0){\rotatebox[origin=t]{90}{\scriptsize flickr-links}} rectangle (4.500,0.976) ;
\draw [fill=\typef,thin] (4.200,0.98) rectangle (4.500,2.039) ;
\draw [fill=\typed,thin] (4.200,2.04) rectangle (4.500,2.195) ;
\draw [fill=\typeb,thin] (4.200,2.20) rectangle (4.500,2.201) ;
\draw [fill=\typeg,thin] (4.700,0) node[below] at (4.85,0){\rotatebox[origin=t]{90}{\scriptsize youtube-links}} rectangle (5.000,3.199) ;
\draw [fill=\typef,thin] (4.700,3.20) rectangle (5.000,3.747) ;
\draw [fill=\typed,thin] (4.700,3.75) rectangle (5.000,3.763) ;
\draw [fill=\typeb,thin] (4.700,3.76) rectangle (5.000,3.763) ;
\draw [fill=\typeg,thin] (5.200,0) node[below] at (5.35,0){\rotatebox[origin=t]{90}{\scriptsize livejournal}} rectangle (5.500,1.118) ;
\draw [fill=\typef,thin] (5.200,1.12) rectangle (5.500,1.990) ;
\draw [fill=\typed,thin] (5.200,1.99) rectangle (5.500,2.084) ;
\draw [fill=\typeb,thin] (5.200,2.08) rectangle (5.500,2.087) ;
\draw [fill=\typeg,thin] (5.700,0) node[below] at (5.85,0){\rotatebox[origin=t]{90}{\scriptsize Average}} rectangle (6.000,1.235) ;
\draw [fill=\typef,thin] (5.700,1.23) rectangle (6.000,1.783) ;
\draw [fill=\typed,thin] (5.700,1.78) rectangle (6.000,1.891) ;
\draw [fill=\typeb,thin] (5.700,1.89) rectangle (6.000,1.902) ;
\draw [<->, thick] (6.20,0) -- (0,0)-- (0,1.80) node[left]{\vspace{10ex}\hspace{-15ex}\rotatebox{90}{Percentage of triangles}} -- (0, 4.00) ;
\draw [dashed] (0, 0.800) node [left]{$20\%$} -- (6.220,0.800);
\draw [dashed] (0, 1.600) node [left]{$40\%$} -- (6.220,1.600);
\draw [dashed] (0, 2.400) node [left]{$60\%$} -- (6.220,2.400);
\draw [dashed] (0, 3.200) node [left]{$80\%$} -- (6.220,3.200);
\draw [dashed] (0, 4.000) node [left]{$100\%$} -- (6.220,4.000);
\draw[fill=\typeg,thin](0.2, 4.2)rectangle(0.5,4.5) node[right] at (0.5,4.35){3-recip};
\draw[fill=\typef,thin](1.7, 4.2)rectangle(2.0,4.5) node[right] at (2.0,4.35){2-recip};
\draw[fill=\typed,thin](3.2, 4.2)rectangle(3.5,4.5) node[right] at (3.5,4.35){1-recip};
\draw[fill=\typeb,thin](4.7, 4.2)rectangle(5.0,4.5) node[right] at (5.0,4.35){simple};
\end{tikzpicture}

%% file: tikz-figures/true-vs-pred.tikz
\begin{tikzpicture} [scale=1.400000e+000]

\draw [thick] [<->] (0,2.3) -- (0,0) -- (4.5,0);
\draw [dashed] (0,0.7) node [left]{10} -- (4.5,0.7) ;
\draw [dashed] (0,1.4) node [left]{100} -- (4.5,1.4) ;
\draw [dashed] (0,2.1) node [left]{1000} -- (4.5,2.1) ;
\node [rotate=90, scale=1] at (-0.700000,1.150000) {True / Predicted};
\node [below] at (0.600000,-0.050000) {\rotatebox[origin=t]{90}{\scriptsize soc-Slashdot0902}};
\node [below] at (1.200000,-0.050000) {\rotatebox[origin=t]{90}{\scriptsize web-Stanford}};
\node [below] at (1.800000,-0.050000) {\rotatebox[origin=t]{90}{\scriptsize web-BerkStan}};
\node [below] at (2.400000,-0.050000) {\rotatebox[origin=t]{90}{\scriptsize web-Google}};
\node [below] at (3.000000,-0.050000) {\rotatebox[origin=t]{90}{\scriptsize soc-Epinions1}};
\node [below] at (3.600000,-0.050000) {\rotatebox[origin=t]{90}{\scriptsize livejournal}};
\draw [fill=\typea,thin] (0.375,0) rectangle (0.525,-0.190187);
\draw [fill=\typee,thin] (0.525,0) rectangle (0.675,0.0649065);
\draw [fill=\typeg,thin] (0.675,0) rectangle (0.825,0.117037);
\draw [fill=\typea,thin] (0.975,0) rectangle (1.125,0.575769);
\draw [fill=\typee,thin] (1.125,0) rectangle (1.275,1.14087);
\draw [fill=\typeg,thin] (1.275,0) rectangle (1.425,2.1);
\draw [fill=\typea,thin] (1.575,0) rectangle (1.725,0.700027);
\draw [fill=\typee,thin] (1.725,0) rectangle (1.875,1.12365);
\draw [fill=\typeg,thin] (1.875,0) rectangle (2.025,2.1);
\draw [fill=\typea,thin] (2.175,0) rectangle (2.325,1.70942);
\draw [fill=\typee,thin] (2.325,0) rectangle (2.475,2.1);
\draw [fill=\typeg,thin] (2.475,0) rectangle (2.625,2.1);
\draw [fill=\typea,thin] (2.775,0) rectangle (2.925,0.090194);
\draw [fill=\typee,thin] (2.925,0) rectangle (3.075,0.200911);
\draw [fill=\typeg,thin] (3.075,0) rectangle (3.225,0.58166);
\draw [fill=\typea,thin] (3.375,0) rectangle (3.525,1.63239);
\draw [fill=\typee,thin] (3.525,0) rectangle (3.675,1.89322);
\draw [fill=\typeg,thin] (3.675,0) rectangle (3.825,2.1);
\draw[fill=\typea,thin] (0.3,2.3) rectangle (0.45,2.45) node[right,scale=1] at (0.5,2.3){$\gcc_{ii,a}$};
\draw[fill=\typee,thin] (1.8,2.3) rectangle (1.95,2.45) node[right,scale=1] at (2,2.3){$\gcc_{iv,e}$};
\draw[fill=\typeg,thin] (3.3,2.3) rectangle (3.45,2.45) node[right,scale=1] at (3.5,2.3){$\gcc_{vi,g}$};
\end{tikzpicture}

%% file: tikz-figures/z-score.tikz
\begin{tikzpicture} [scale=1.400000e+000]

\draw [thick] [<->] (0,2.3) -- (0,0) -- (4.5,0);
\draw [dashed] (0,0.7) node [left]{10} -- (4.5,0.7) ;
\draw [dashed] (0,1.4) node [left]{100} -- (4.5,1.4) ;
\draw [dashed] (0,2.1) node [left]{1000} -- (4.5,2.1) ;
\node [rotate=90, scale=1] at (-0.700000,1.150000) {Z-score};
\node [below] at (0.600000,-0.050000) {\rotatebox[origin=t]{90}{\scriptsize soc-Slashdot0902}};
\node [below] at (1.200000,-0.050000) {\rotatebox[origin=t]{90}{\scriptsize web-Stanford}};
\node [below] at (1.800000,-0.050000) {\rotatebox[origin=t]{90}{\scriptsize web-BerkStan}};
\node [below] at (2.400000,-0.050000) {\rotatebox[origin=t]{90}{\scriptsize web-Google}};
\node [below] at (3.000000,-0.050000) {\rotatebox[origin=t]{90}{\scriptsize soc-Epinions1}};
\node [below] at (3.600000,-0.050000) {\rotatebox[origin=t]{90}{\scriptsize livejournal}};
\draw [fill=\typea,thin] (0.375,0) rectangle (0.525,-0.826351);
\draw [fill=\typee,thin] (0.525,0) rectangle (0.675,0.97306);
\draw [fill=\typeg,thin] (0.675,0) rectangle (0.825,1.33073);
\draw [fill=\typea,thin] (0.975,0) rectangle (1.125,1.67255);
\draw [fill=\typee,thin] (1.125,0) rectangle (1.275,2.1);
\draw [fill=\typeg,thin] (1.275,0) rectangle (1.425,2.1);
\draw [fill=\typea,thin] (1.575,0) rectangle (1.725,1.89054);
\draw [fill=\typee,thin] (1.725,0) rectangle (1.875,2.1);
\draw [fill=\typeg,thin] (1.875,0) rectangle (2.025,2.1);
\draw [fill=\typea,thin] (2.175,0) rectangle (2.325,2.1);
\draw [fill=\typee,thin] (2.325,0) rectangle (2.475,2.1);
\draw [fill=\typeg,thin] (2.475,0) rectangle (2.625,2.1);
\draw [fill=\typea,thin] (2.775,0) rectangle (2.925,1.4523);
\draw [fill=\typee,thin] (2.925,0) rectangle (3.075,1.40087);
\draw [fill=\typeg,thin] (3.075,0) rectangle (3.225,1.84777);
\draw [fill=\typea,thin] (3.375,0) rectangle (3.525,2.1);
\draw [fill=\typee,thin] (3.525,0) rectangle (3.675,2.1);
\draw [fill=\typeg,thin] (3.675,0) rectangle (3.825,2.1);
\draw[fill=\typea,thin] (0.3,2.3) rectangle (0.45,2.45) node[right,scale=1] at (0.5,2.3){$\gcc_{ii,a}$};
\draw[fill=\typee,thin] (1.8,2.3) rectangle (1.95,2.45) node[right,scale=1] at (2,2.3){$\gcc_{iv,e}$};
\draw[fill=\typeg,thin] (3.3,2.3) rectangle (3.45,2.45) node[right,scale=1] at (3.5,2.3){$\gcc_{vi,g}$};
\end{tikzpicture}